\documentclass[preprint,11pt]{elsarticle}
\usepackage[
top=1.8cm,
bottom=2.0cm,
left=1.8cm,
right=1.8cm
]{geometry}
\biboptions{sort&compress}
\usepackage{graphicx}
\usepackage{longtable,booktabs}
\usepackage{amsmath}
\usepackage{amssymb}
\usepackage{bm}
\usepackage{placeins}
\usepackage{color}
\usepackage{subcaption}
\usepackage[pdftex,breaklinks,colorlinks,
citecolor=blue,
urlcolor=blue]{hyperref}
\usepackage{microtype}
\microtypesetup{expansion=false}

\begin{document}

\begin{frontmatter}

\title{Deconvoluting Mechanisms of Radiation-Induced Grain Boundary Segregation in Dilute Ni--X (X = Fe, Cr, Mn) Alloys via Ion Irradiation and Phase-field Modeling}

\author[1]{Sohail Shah\corref{cor1}\fnref{eq}}
\author[2]{Sourabh Bhagwan Kadambi\corref{cor1}\fnref{eq}}
\author[1]{Anshul Kamboj}
\author[4]{Benjamin Diaz}
\author[1]{Xinchang Zhang}
\author[1]{Jana Howard}
\author[3]{Michael Moorehead}
\author[1]{Boopathy Kombaiah}
\author[4]{Lin Shao}
\author[1]{Mukesh Bachhav}

\cortext[cor1]{Corresponding authors: Sohail.Shah@inl.gov (S. Shah), SourabhBhagwan.Kadambi@inl.gov (S. B. Kadambi).}
\fntext[eq]{These authors contributed equally to the writing of the manuscript.}
 
\affiliation[1]{organization={Materials and Fuels Complex, Idaho National Laboratory},
                city={Idaho Falls},
                state={ID},
                postcode={83415},
                country={USA}}
\affiliation[2]{organization={Computational Mechanics and Materials Department, Idaho National Laboratory},
                city={Idaho Falls},
                state={ID},
                postcode={83415},
                country={USA}}
\affiliation[3]{organization={Irradiated Fuels and Materials Department, Idaho National Laboratory},
                city={Idaho Falls},
                state={ID},
                postcode={83415},
                country={USA}}
\affiliation[4]{organization={Department of Nuclear Engineering, Texas A\&M University},
                city={College Station},
                state={TX},
                postcode={77845},
                country={USA}}

\begin{abstract}
Radiation-induced composition redistribution at grain boundaries (GBs) in Ni-based structural alloys poses a critical challenge for the long-term deployment of advanced nuclear reactors.
However, the mechanisms governing solute-specific segregation remain poorly understood because defect-mediated transport, thermodynamic driving forces, and competing microstructural sinks can all contribute under irradiation.
In this work, we perform systematic Ni$^{2+}$ ion irradiations on dilute Ni--Fe, Ni--Cr, and Ni--Mn binary alloys at 500, 800, and 1000~K to 2~dpa and characterize GB chemistry using scanning transmission electron microscopy-energy-dispersive X-ray spectroscopy.
The segregation behavior is then interpreted using rate-theory phase-field modeling informed by atomistic transport data.
We find that Fe depletes at GBs across all temperatures through radiation-induced segregation (RIS) via preferential vacancy-solute exchange mechanism (inverse Kirkendall effect).
Cr segregation, on the other hand, reflects a temperature-dependent RIS with competition between self-interstitial atom (SIA)-mediated transport toward GBs and vacancy-mediated transport away from GBs, yielding near-zero net segregation at 500~K and depletion at higher temperatures.
In contrast to Fe and Cr, Mn exhibits a distinct behavior dominated by thermodynamic segregation that drives local enrichment at the GB.
We leverage depth-dependent characterization of segregation along the same GB, spanning irradiated and unirradiated regions, to separate radiation-induced and thermodynamic contributions, and validate the unirradiated region as an internal thermodynamic reference.
Void formation within the grains at 800~K and irradiation-induced GB migration at 800 and 1000~K is found to further add mechanistic complexity by suppressing and redistributing the solute. Our combined experimental and computational assessments show that the interpretation of GB chemistry evolution under irradiation must consider concurrent contributions from non-equilibrium kinetic transport, thermodynamic driving forces, bulk sink effects, and GB migration.
\end{abstract}


\end{frontmatter}

\section{Introduction}
\label{sec:intro}

Ni-based alloys have emerged as promising structural material candidates for certain advanced nuclear reactor concepts, particularly high-temperature gas-cooled and molten salt reactors, owing to their high-temperature mechanical strength, creep resistance, and corrosion tolerance in relevant service environments \cite{rowcliffe2009perspectives,nickel1984evaluation}.
Under prolonged exposure to irradiation, however, these alloys develop supersaturated point defect concentrations that drive local chemistry changes at extended defects such as GBs and dislocations.
Such compositional changes can alter phase stability, promote embrittlement and intergranular degradation, and reduce corrosion resistance, thereby limiting reactor licensing and long-term deployment~\cite{was2007fundamentals,zinkle2013materials}.
Consequently, understanding the mechanisms governing solute redistribution is essential for controlling intergranular phase formation, coupled irradiation–corrosion effects, and for informing alloy design for deployment viability.

Radiation-induced composition redistribution of alloying elements at point defect sinks has traditionally been interpreted as arising from the coupling between non-equilibrium point defect fluxes and atomic transport under irradiation~\cite{was2007fundamentals,ardell2016,nastar2012}.
Numerous experimental and modeling studies have reported solute segregation at GBs in Ni-based and Fe-based alloys~\cite{barr2015, briggs2016,allen1998acta}, often attributing such behavior to the non-equilibrium defect-coupled transport processes \cite{was2007fundamentals,zinkle2013materials}.
For example, preferential vacancy--solute exchange drives solute depletion at sinks when the solute diffuses faster than the solvent (inverse Kirkendall (IK) effect), whereas preferential SIA-mediated transport drives solute enrichment at sinks through solute--dumbbell drag~\cite{wiedersich1979,was2007fundamentals}, with the net segregation determined by the competition between these two mechanisms.
Such interpretations are typically framed within a purely kinetic framework, collectively referred to as RIS~\cite{song1998grain,wiedersich1979}, which neglects potential contributions from equilibrium thermodynamic driving forces.
Although thermodynamic contributions to GB segregation under irradiation were proposed in early theoretical treatments~\cite{faulkner1996model,nastar2005}, direct experimental evidence in Ni-based alloys remains scarce.
Advances in high-resolution elemental characterization techniques have enabled direct observation of these contributions in Fe-based alloys~\cite{barr2018oscillatory_ris,lach2021stem_apt_ris,thomas2025decoupling}.
Recent phase-field modeling has demonstrated that thermodynamic GB segregation (TS) of Cr and Ni in fcc Fe--Cr--Ni can be opposite in sign to RIS~\cite{kadambi2025}, providing a mechanistic basis for experimentally observed non-monotonic (e.g., W-shaped or oscillatory) segregation profiles~\cite{barr2018oscillatory_ris,lach2021stem_apt_ris} that cannot be explained by RIS alone.
Similarly, recent experimental and kinetic Monte Carlo studies have shown that comparable peak GB segregation can arise under both irradiated and non-irradiated conditions, while irradiation primarily broadens the segregation profile \cite{thomas2025decoupling}.
These observations indicate that segregation behavior cannot be interpreted solely within a non-equilibrium RIS framework, highlighting the need to decouple equilibrium thermodynamic and non-equilibrium kinetic contributions to solute redistribution.
These findings motivate a distinction between RIS, arising from non-equilibrium defect-driven transport, and radiation-enhanced thermodynamic segregation (RETS), which reflects the thermodynamic preference of solutes for GBs with kinetics accelerated under irradiation.

Traditional RIS measurements on commercial alloys are intrinsically complicated because multiple alloying elements and impurities can segregate simultaneously and also partition into irradiation-induced clusters and precipitates \cite{lach2021stem_apt_ris,barr2018oscillatory_ris}.
In multicomponent Ni-based alloys, this coupled redistribution obscures the intrinsic segregation tendencies of individual solutes, making it difficult to isolate and quantify RIS and RETS contributions independently~\cite{was2007fundamentals,ardell2016}.
Interpretation is further complicated by competing defect sinks, such as voids and dislocation loops, and by irradiation-induced GB migration, both of which modify the segregation state independently of solute transport \cite{brailsford1981,barr2022science}.
To isolate these contributions, systematic study of segregation in well-controlled model alloys with simplified compositions is essential for decoupling RIS and RETS from confounding microstructural factors.

Mechanistic models have focused predominantly on defect-mediated RIS, with several ab initio-informed rate-theory studies using density functional theory (DFT) energetics and migration barriers to parameterize vacancy- and SIA-mediated diffusion in Ni--Cr alloys \cite{barnard2012,barnard2014aimd,tucker2010}.
However, SIA-mediated transport in fcc Ni alloys remains difficult to quantify even within atomistic modeling frameworks due to the complexity of SIA configurations, including mixed-dumbbell formation, multiple migration pathways, and strong solute--defect interactions, which are not directly accessible experimentally \cite{barnard2014aimd,tucker2010}.
These challenges are even more severe in concentrated alloys, where multisolute interactions further complicate the calculation of transport coefficients, so most atomistically informed studies have focused on dilute alloy limits~\cite{toijer_solute-point_2021,shousha2024vacancy}.
Even here, different modeling approaches have led to contradictory predictions regarding the role of SIA transport, with some studies indicating that SIA fluxes can promote Cr enrichment at GBs, while others predict vacancy-dominated behavior leading to Cr depletion, highlighting the sensitivity of RIS predictions to uncertainties in SIA transport coefficients \cite{barnard2012,barnard2014aimd}.  
More recently, SCMF theory combined with DFT has enabled direct calculation of multicomponent transport coefficients, though currently feasible primarily in the dilute-alloy limit~\cite{toijer_solute-point_2021,shousha2024vacancy}.
Notably, Toijer et al.~\cite{toijer_solute-point_2021} predicted Fe and Mn depletion via vacancy-dominated transport, but predicted Cr to exhibit a temperature-dependent crossover from enrichment (SIA-dominated transport) at low temperature to depletion (via vacancy-dominated transport) at high temperature.
Experimental studies on dilute alloy behaviour are therefore essential to isolate the intrinsic segregation tendencies of individual solutes in fcc Ni and to compare against existing predictions \cite{toijer_solute-point_2021,shousha2024vacancy}.
Resolving these competing contributions requires an integrated approach combining nanoscale experimental characterization \cite{shah2024correlating,curtis2025local,shah2025irradiation,shah2026electric} with mechanistic modeling \cite{oplinger2026,kadambi2025}.

In this study, we employ dilute Ni--X (X = Fe, Cr, Mn) binary alloys to systematically investigate the mechanisms governing solute segregation under irradiation.
Specifically, we examine how the relative contributions of vacancy- and SIA-mediated transport control element-specific segregation, how changes in the dominant transport mechanism produce temperature-dependent segregation, and whether thermodynamic driving forces contribute alongside RIS.
We further assess how competing bulk sinks and GB migration modify the resulting segregation behavior.
Ion irradiations using Ni$^{2+}$ are performed on high-purity alloys with large grain sizes and minimal pre-existing dislocations or precipitates, to minimize confounding microstructural contributions to the measured segregation signal.
Segregation is systematically investigated under pre-irradiation thermal annealing and across ion depths encompassing both irradiated and unirradiated reference regions, at temperatures of 500, 800, and 1000~K and doses up to $\sim$2~dpa.
The experimental observations are interpreted using a rate-theory phase-field modeling framework parameterized by atomistically-informed properties for both vacancy- and SIA-mediated diffusion~\cite{toijer_solute-point_2021,oplinger2026}.
Our modeling framework introduces a coupled treatment of RIS and RETS, two potentially competing segregation mechanisms that are rarely distinguished in prior models, together with GB migration, using atomistically informed vacancy- and SIA-mediated Onsager transport within a single rate-theory phase-field framework. The framework further extends this treatment to two dimensions, capturing the full depth-dependent ion-damage field rather than a single depth-averaged condition.
This integrated approach provides a direct basis for reconciling experimental observations with mechanistic predictions for individual solutes in fcc Ni.

\section{Methodology} \label{sec:methods}

\subsection{Experimental} \label{sec:exp_methods}

Dilute alloys of composition Ni--1Fe (Ni -- 1~at.\% Fe), Ni--1Cr (Ni -- 1~at.\% Cr), and Ni--1Mn (Ni -- 1~at.\% Mn) were investigated in this study. Raw elemental materials, each in the form of irregular pieces with purities $>$99.9\% (metals basis), were sourced from Alfa Aesar (Thermo Fisher Scientific). 10-g charges of Ni-1Fe, Ni-1Cr, and Ni-1Mn (at.\%) were weighed out using a Mettler Toledo XS204 analytical balance and subsequently arc melted into individual ingots, which were flipped and remelted multiple times to ensure chemical homogeneity, using a Centorr 5SA series single-arc furnace. Arc-melted samples were then encapsulated in quartz tubes under vacuum, heated to 1473 K (1200~\textdegree C) at a rate of 5~\textdegree C/min, then held at that temperature for 48 hours before water quenching to homogenize any chemical microsegregation originating from solidification during arc melting. All heat treatments were performed under identical vacuum-sealed conditions.
Self-ion irradiations were then performed using a 3 MV National Electrostatics Corporation (NEC) tandem accelerator at Texas A\&M University.
A beam current of 160 nA was used with a defocused 5 MeV Ni$^{2+}$ ion beam (no wobbling or rastering) and a beam spot size of 6.5~mm $\times$ 6.5~mm focused on the target.
The damage profile (see Fig.~S5 in the Supplementary Material) was estimated using the Stopping and Range of Ions in Matter (SRIM) code in Kinchin--Pease mode \cite{ziegler2010srim,stoller2013use}.
Irradiations were conducted to a peak dose of 2 displacements per atom (dpa) at a depth of $\sim$1.5~$\mu$m, corresponding to a peak damage rate of $1.5\times10^{-3}$ dpa/s.
The fluence was $2.04 \times 10^{15}$ ions/cm$^2$ with a flux of $1.54 \times 10^{12}$ ions/cm$^2$/s.
During irradiation, the chamber vacuum was maintained at $6 \times 10^{-8}$ torr or lower.
A liquid nitrogen trap was positioned near the target assembly to reduce contamination and improve the local vacuum.
Irradiations were performed at 500~K ($\sim$230~\textdegree C), 800~K($\sim$530~\textdegree C), and 1000~K($\sim$730~\textdegree C), with temperature monitored using a thermocouple mounted near the irradiated region.  These temperatures capture the transition from limited defect mobility and damage accumulation to enhanced diffusion, defect recombination, segregation, and high-temperature microstructural recovery/stability.

Electron-transparent specimens for GB segregation analysis were prepared using a dual-beam focused ion beam/scanning electron microscope (FIB/SEM, FEI Quanta 3D FEG).
FIB lamellae were extracted across high-angle GBs, which serve as effective sinks for irradiation-produced point defects. Electron backscatter diffraction (EBSD) analysis was performed to characterize the crystallographic orientation distribution and identify high-angle GBs. Orientation maps were acquired from the polished specimen surface and indexed using standard EBSD processing routines. GBs were classified according to the crystallographic misorientation between adjacent pixels/grains, with boundaries exhibiting misorientation angles greater than 15° defined as high-angle GBs. These boundaries were subsequently extracted and used for quantitative microstructural analysis.
Final thinning was carried out at 5 kV and 48 pA, followed by a 2 kV cleaning step.
Typical transmission electron microscopy (TEM) lamella dimensions were approximately $15 \times 10$~$\mu$m.

GB segregation was characterized using a Thermo Fisher Spectra 300 monochromated, double-aberration-corrected 300 kV scanning transmission electron microscope equipped with a Super-X energy-dispersive X-ray spectroscopy (EDS) detector.
EDS maps were acquired simultaneously with high-angle annular dark-field (HAADF) STEM imaging for spatial correlation, using identical acquisition parameters to enable qualitative comparison between regions.
Quantitative analysis of the EDS data was performed using Velox software. The measured matrix compositions were consistently ~0.5–1 at.\% higher than the nominal alloy compositions. Such deviations are not uncommon in quantitative STEM-EDS measurements of dilute alloy systems and can arise from uncertainties associated with X-ray generation volume, specimen thickness variations, detector geometry, background subtraction, absorption corrections, and counting statistics, particularly when quantifying solute concentrations near the detection limit ~\cite{goldstein2017scanning,cliff1975quantitative}. Previous studies have similarly reported that quantitative EDS measurements in TEM foils are most reliable for assessing relative compositional variations rather than absolute solute concentrations in dilute alloys. Therefore, segregation magnitudes reported in this work are referenced to the local matrix composition adjacent to each GB rather than the nominal alloy composition, ensuring that the interpretation of enrichment or depletion remains unaffected by systematic offsets in absolute quantification. 

Segregation quantification was performed using \(120 \times 50~\mathrm{nm}^2\) region of interest (ROI) in the bulk and \(120 \times 3~\mathrm{nm}^2\) spectrum-integration ROI across the GB. The error bars were then calculated from three such values obtained at distinct locations within each map. A total of 13 independent random high-angle GBs were characterized across the nine alloy--temperature conditions, with three independent boundaries analyzed for Ni--1Cr at 800 and 1000~K to assess boundary-to-boundary reproducibility (Supplementary Material, Sec.~S9); one boundary was analyzed for each of the remaining eight conditions.
To improve signal-to-noise ratio in the dilute-alloy EDS datasets, elemental maps were post-processed in Thermo Fisher Velox using an average (mean over three pixels) spatial filter prior to quantitative profile extraction. Spatial averaging is commonly employed in STEM-EDS analysis to suppress Poisson counting noise and improve visualization of low-concentration elemental distributions while preserving composition gradients that extend over length scales substantially larger than the filter kernel \cite{carter2009transmission,kotula2003automated}. Following filtering, composition profiles were extracted using identical procedures for all datasets. Because the same filtering parameters were applied consistently to all measurements, the filtering affected only the statistical noise level and did not alter the observed segregation trends or the relative enrichment/depletion behavior at GBs.

\subsection{Modeling} \label{sec:model_methods}

\subsection{Governing equations}
\label{sec:governing_equations}
To describe the coupled evolution of solute redistribution, thermodynamic segregation (TS), and GB migration in a bicrystal geometry, we employ the rate-theory phase-field framework developed in our previous work of Kadambi et al.~\cite{kadambi2025}, and apply it to dilute Ni--X binary alloys.
Concentrations in the bulk and GB regions evolve under the competing action of RIS, driven by non-equilibrium defect fluxes, and TS, driven by the solute's chemical affinity for the GB.
The governing equations track the evolution of the solute concentration, vacancies, and SIAs in the lattice frame of reference (with the solvent treated implicitly):
\begin{subequations}
\begin{flalign}
    &\frac{\partial c_{X}}{\partial t} = -\nabla \cdot \boldsymbol{J}_{X}, \\
    &\frac{\partial c_{V}}{\partial t} = -\nabla \cdot \boldsymbol{J}_{V} + P_{V} 
    - R_{VI}\, c_Vc_I 
    - k^2_b D_{V}(c_{V}-c^e_{V}) 
    - k^2_g D_{V}(c_{V}-c^e_{V})\, g_{\text{sink}}(\eta), \\
    &\frac{\partial c_{I}}{\partial t} = -\nabla \cdot \boldsymbol{J}_{I} + P_{I} 
    - R_{VI}\, c_Vc_I 
    - k^2_b D_{I} c_{I}
    - k^2_g D_{I} c_{I}\, g_{\text{sink}}(\eta). &&
\end{flalign}
\end{subequations}
Here, $k^2_g$ is the local GB sink strength and $g_{\text{sink}} = 65536\,\eta^8(1-\eta)^8$ is a narrow bell-shaped sink indicator function~\cite{kadambi2025} that equals unity at the GB center ($\eta = 0.5$) and vanishes in the grain interiors.
The GB sink strength is defined as $k^2_g = 2\alpha^{-1}/(\delta C_{\text{sink}})$, where $\alpha^{-1}$ governs the rate of point defect absorption by the GB and can range from non-ideal to ideal-sink behavior depending on GB structure~\cite{duh2001numerical,gu2017point}$,$ $\delta$ is the GB width, and $C_{\text{sink}} = 0.32$ is the integration constant specific to the chosen diffuse sink function $g_{\text{sink}}$~\cite{kadambi2025}. $\boldsymbol{J}$ denotes the flux vector, $P_V$ and $P_I$ are the vacancy and SIA production rates (assumed equal to isolate the role of transport coefficients on segregation behavior, independent of defect production asymmetry~\cite{oplinger2026}), and $R_{VI} = 4\pi r_\circ(D_V+D_I)/V_a$ is the vacancy--SIA recombination rate, where $r_\circ$ is the recombination radius and $V_a$ is the atomic site volume. $D_V$ and $D_I$ are the vacancy and SIA diffusivities, respectively.
The bulk sink strength $k^2_b$ describes an effective point defect absorption by mean-field vacancy clusters, voids, SIA loops, and dislocation networks.
For simplicity, we assume equal sink strengths for absorption of vacancies and SIAs. Absorption bias can readily be introduced into the framework~\cite{kadambi2025,oplinger2026} and can be considered in future work to more rigorously treat the sink behavior of dislocations and SIA loops.
The coupled fluxes of atoms and point defects are written using linear irreversible thermodynamics as
\begin{subequations}
\begin{flalign}
    &\boldsymbol{J}_{X} = - \tilde{L}_{XX} \nabla \tilde{\mu}_{X} - L_{XV} \nabla \mu_{V} - L_{XI} \nabla \mu_{I}, \\
    &\boldsymbol{J}_{V} = - \tilde{L}_{VX} \nabla \tilde{\mu}_{X} - L_{VV} \nabla \mu_{V}, \\
    &\boldsymbol{J}_{I} = - \tilde{L}_{IX} \nabla \tilde{\mu}_{X} - L_{II} \nabla \mu_{I}, &&
\end{flalign}
\label{eq:total_flux_1}
\end{subequations}
where lattice site conservation under unbiased conditions requires $\boldsymbol{J}_{Ni} + \boldsymbol{J}_{X} + \boldsymbol{J}_{V} - \boldsymbol{J}_I = 0$.
The diffusion potential of solute $X$ relative to solvent Ni is $\tilde{\mu}_{X} = \mu_{X} - \mu_{Ni} = \partial f_{\textrm{alloy}}(c_{X})/\partial c_{X}$, where $f_\textrm{alloy}(c_{X})$ is the alloy free-energy density.
The point-defect chemical potentials under ideal solution assumption are given by $\mu_{V}=(RT/V_m)\ln(c_V/c^e_V)$ and $\mu_{I}=(RT/V_m)\ln(c_I/c^e_I)$, where $c^e$ is the thermal equilibrium point-defect concentration and $V_m$ is the constant molar lattice volume.
The relative transport coefficients are expressed in terms of the Onsager coefficients as
\begin{subequations}
\begin{flalign}
    &\tilde{L}_{VX} = - L^{V}_{XX} - L^{V}_{XNi} + c_{X} L_{VV},\\
    &\tilde{L}_{IX} = L^{I}_{XX} + L^{I}_{XNi} - c_{X} L_{II}, \\
    &\tilde{L}_{XX} = L^{V}_{XX} + L^{I}_{XX} + \, c_{X} L_{XV} - \, c_{X} L_{XI}. &&
\end{flalign}
\label{eq:transport_coeff}
\end{subequations}
Assuming that point defect concentrations remain in the dilute limit (below $10^{-3}$), we take $c_{Ni}=1-c_{X}$ in the thermodynamic description of the alloy.
We further note that $D_V = L_{VV}\theta_{VV}$ and $D_I = L_{II}\theta_{II}$.

The isothermal grand-potential functional for the bicrystal microstructure can be written in terms of a single phase-field order parameter $\eta$ as
  \begin{flalign} \label{eq:GP_functional}
      {\Omega}
      = \int_V
      \left[\omega^b + \left(m_0 + \omega^g-\omega^b\right)\bar{g}(\eta) + \Delta f\,h(\eta) + \frac{\kappa}{2} \left|\nabla \eta \right|^2 \right]
   d V, &&
  \end{flalign}
\noindent where the grand-potential densities for the bulk ($b$) and GB ($g$) are given by $\omega^{b, g} = f^{b, g} - c^{b, g}_{X} \tilde{\mu}_{X} - c_{V} \mu_{V}  - c_{I} \mu_{I}$.
The phase concentrations $c^{b}_{X}$ and $c^{g}_{X}$ are obtained from the respective free-energy densities via the diffusion potential, while point defect concentrations $c_V$ and $c_I$ do not carry phase dependence and are instead related to their chemical potentials through a single thermodynamic expression.
The explicit phase-concentration and point-defect relations are given in Sec.~S1.
$\Delta f$ represents a prescribed bulk free-energy density difference between the two grains used to drive GB migration, and $h(\eta)$ is the switching function through which $\Delta f$ is applied to one of the two grains.
Here, $\omega^g-\omega^b$ is the excess grand-potential density in the GB region, $\bar{g}=16\eta^2(1-\eta)^2$ is the normalized double-well potential, $m_0$ is a barrier height capturing the pure-component (Ni) grand-potential difference, and $\kappa$ is the gradient-energy coefficient for the order-parameter fields.
For simplicity, distinct GB energetics are included only for the alloy component, while distinct point-defect energetics at the GB are neglected.
Gradient-energy contributions are included for the order-parameter fields but neglected for the concentration fields, per standard practice in phase-field modeling.
Assuming ideal-solution behavior, the free-energy density of the alloy with point defects is given by
\begin{subequations}
\begin{flalign} 
        &f^b = \frac{RT}{V_m} \left[c_X \ln{c_X} + c_V \ln{\frac{c_V}{c^e_V}} + c_I \ln{\frac{c_I}{c^e_I}} \right], \\ % E^f_V c_V + E^f_I c_I +
    &f^g = f^b + \Delta E^\textrm{seg}_X c_X, &&
\end{flalign} 
\label{eq:fe_taylor}
\end{subequations}
\noindent where $f^b$ is the bulk free-energy density expressed under the dilute-solution approximation and $f^g$ is the GB free-energy density.
The order parameter $\eta$ serves as a grain indicator field, with $\eta = 1$ in one grain and $\eta = 0$ in the adjacent grain, and transitions smoothly across the diffuse GB region.
The segregation energy $\Delta E^\textrm{seg}_X$ quantifies the thermodynamic preference of solute $X$ for the GB relative to the bulk; this term is responsible for modeling both equilibrium TS and its radiation-enhanced counterpart (RETS), since irradiation accelerates the kinetics to achieve the equilibrium set by $\Delta E^\textrm{seg}_X$.
The corresponding time-evolution equation for the bicrystal order parameter introduces the driving force for GB migration:
\begin{flalign} \label{eq:eta_evol}
\frac{\partial \eta}{\partial t} = -L_\eta\, \frac{\delta \Omega}{\delta \eta} = -L_\eta\, \left[\left(m_0 + \omega^g-\omega^b\right)\frac{d\bar{g}}{d\eta} + \Delta f\,\frac{dh}{d\eta} - \kappa\nabla^2\eta \right], &&
\end{flalign}
A nonzero $\Delta f$ introduces an asymmetry between the two grains, causing the GB to migrate toward the higher-energy grain; this mechanism is used in the moving-GB simulations of Sec.~\ref{subsec:model_results:moving_GB}.

\subsection{Parameterization}
\label{sec:parameterization}

Following the Manning-type formulation of atomic and point-defect transport coefficients used for concentrated alloys~\cite{manning1971,lidiard1986,yang2016roles,kadambi2025}, the Onsager coefficients in Eq.~\ref{eq:manning_binary} were derived from the vacancy- and SIA-mediated partial diffusivities $d_{k\nu}$ as:
\begin{subequations}
\begin{flalign}
L^{\nu}_{kl} = \frac{c_k c_{\nu} d_{k\nu}}{RT/V_m} \left(\delta_{kl} + \frac{2 c_l d_{l\nu}}{M_0 \left(c_{Ni}d_{Ni\nu}+c_X d_{X\nu}\right)} \right), &&
\label{eq:manning_binary}
\end{flalign}
\begin{flalign}
d_{kI} = \frac{1}{6}\lambda_I^2 z_I f_I \omega^\circ_{kI}\exp\!\left(-\frac{E_m^I}{k_B T}\right)\beta_{kI}, &&
\label{eq:sia_partial_diff}
\end{flalign}
\end{subequations}
where $k,l=\mathrm{Ni},X$, $\nu=V$ or $I$, $\delta_{kl}$ is the Kronecker delta, $M_0=2f_0/(1-f_0)$, and $c_{Ni}=1-c_X$. Here, $d_{k\nu}$ denotes the partial diffusivity of species $k$ mediated by point defect $\nu$.
Vacancy partial diffusivities were parameterized using the temperature-dependent data of Toijer et al.~\cite{toijer_solute-point_2021}.
These data provide a self-consistent set of vacancy-mediated transport coefficients for Ni, Fe, Cr, and Mn in fcc Ni derived from DFT migration barriers and SCMF theory as implemented in KINECLUE.
To correct for the systematic underestimation of vacancy formation energies by generalized gradient approximation density functional theory (GGA-DFT), Toijer et al. treated the equilibrium vacancy concentration as a fitting parameter based on experimental self-diffusion data, yielding $c^e_V = \exp\left(-(1.8 - 4k_BT)/k_BT\right)$.
Arrhenius fits to the resulting diffusivity data of Ni, Fe and Cr (see Fig.~S1 in Supplementary Material) were used directly in the rate-theory phase-field model.
For Mn, the radiation-enhanced diffusion coefficient reported by Toijer et al. (computed at $c_V = 10^{-6}$) was reduced to the thermal diffusion coefficient using the fitted equilibrium vacancy concentration (see Fig.~S3).
The resulting vacancy partial diffusivities follow the hierarchy $d_{\mathrm{Fe}V} > d_{\mathrm{Mn}V} > d_{\mathrm{Cr}V} > d_{\mathrm{Ni}V}$.

\begin{figure}[htp!]
\centering
    \begin{subfigure}[t]{0.48\textwidth}
        \includegraphics[width=1\textwidth]{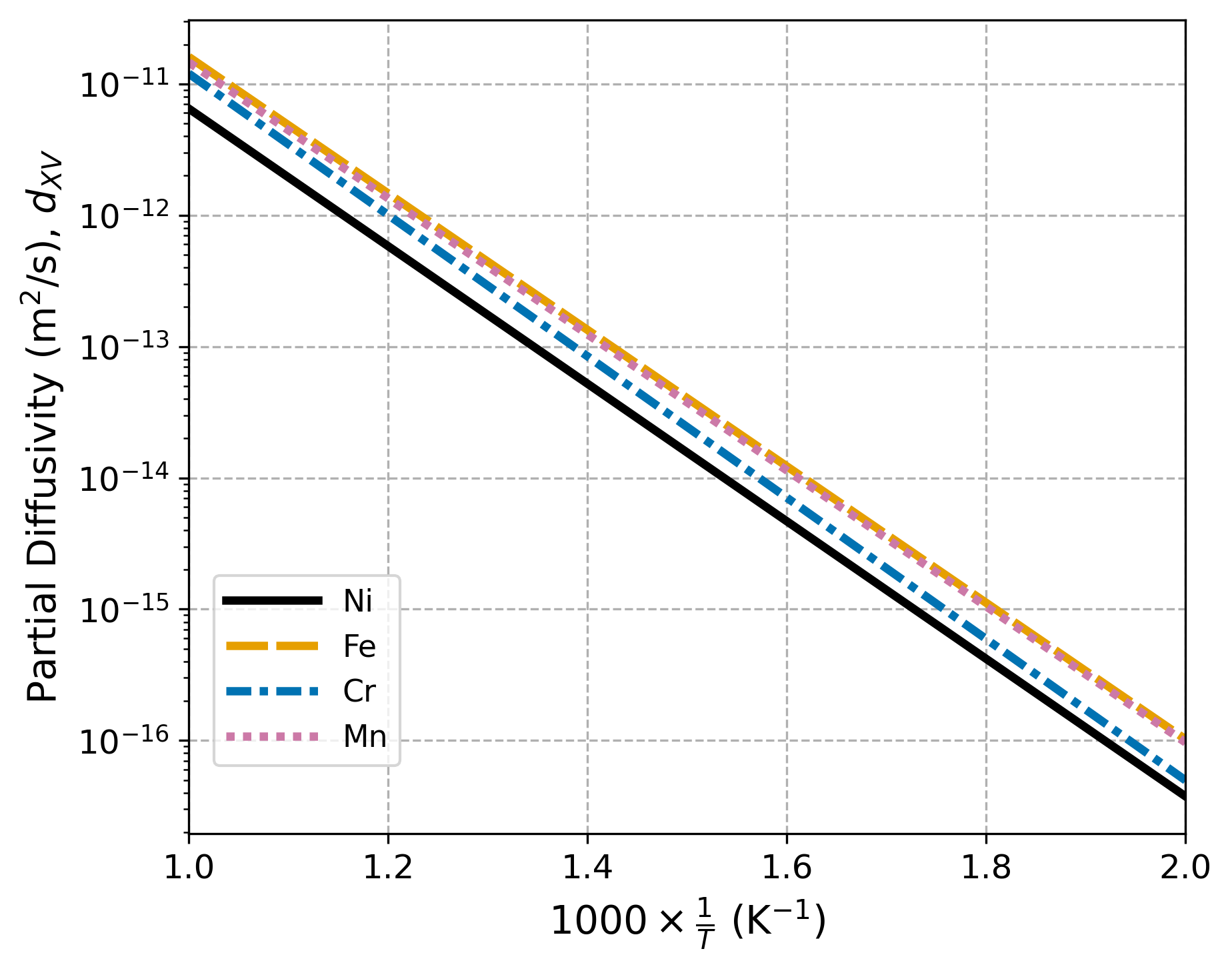}
    \caption{}
    \label{subfig:diff_data_diff}
    \end{subfigure}
    \begin{subfigure}[t]{0.485\textwidth}
        \includegraphics[width=1\textwidth]{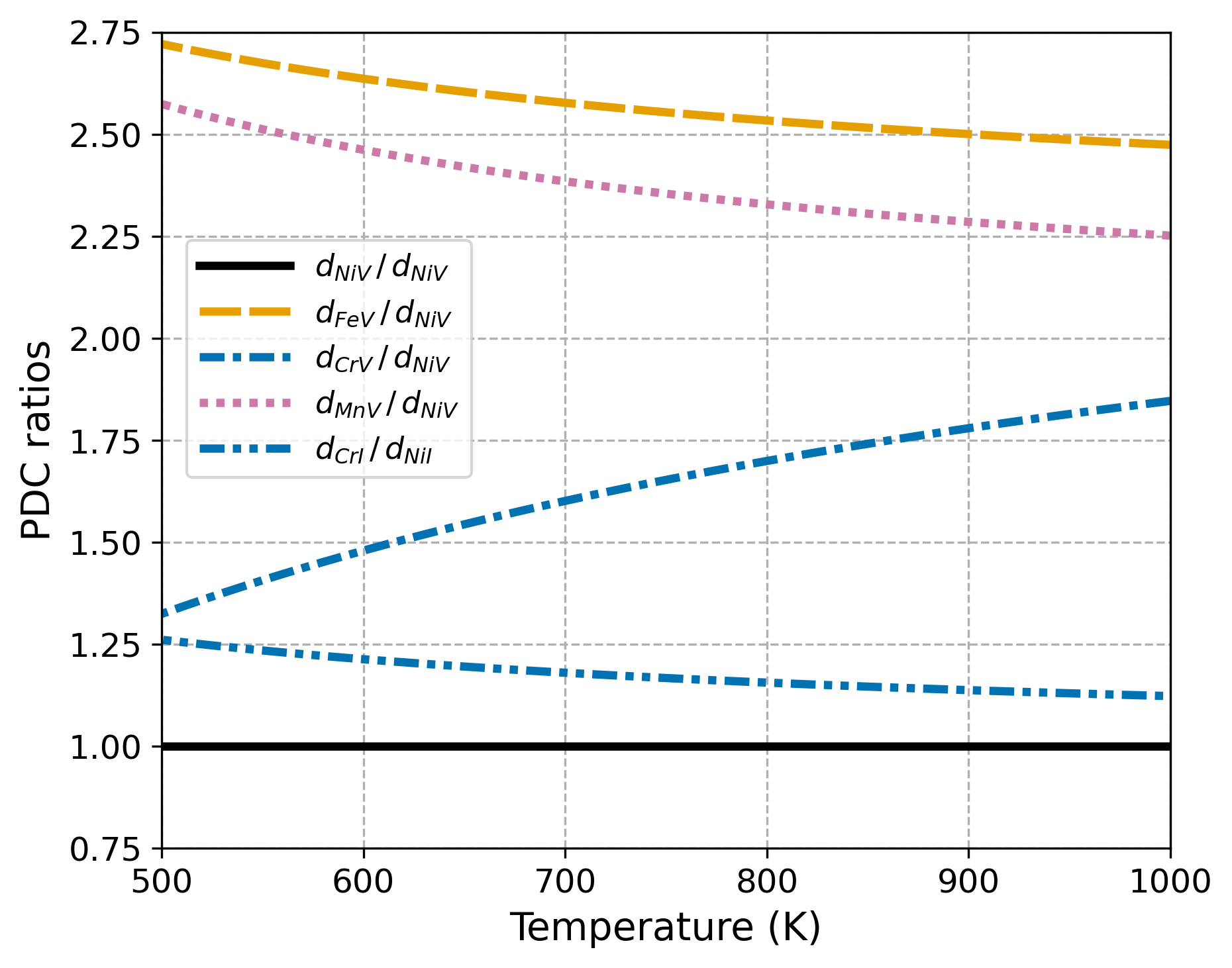}
    \caption{}
    \label{subfig:diff_data_pdc}
    \end{subfigure}
    \caption{(a) Vacancy partial diffusivities of Ni, Fe, Cr, and Mn as a function of inverse temperature as derived from Toijer et al.~\cite{toijer_solute-point_2021}.
    (b) Partial diffusion coefficient (PDC) ratios for vacancy transport of solutes relative to Ni.
    PDC ratio for Cr-SIA transport estimated from Cr-SIA binding energy.}
    \label{fig:diff_data}
\end{figure}

SIA partial diffusivities were parameterized using Eq.~\ref{eq:sia_partial_diff}, with a migration energy of 0.6 eV. The prefactor in Eq.~\ref{eq:sia_partial_diff} captures the geometric, correlation, and attempt-frequency terms; parameter definitions and their values are provided in Sec.~S3 of the Supplementary Material. The equilibrium SIA concentration is obtained at the formation energy $E_f^I = 2.0$~eV.${\beta}_{kI}=\exp{\left((E_{kI}-E_{NiI})/k_BT)\right)} / \left\{c_{Ni}+c_{k}\exp{\left((E_{kI}-E_{NiI})/k_BT\right)}\right\}$ is the relative solute--SIA binding factor, with $E_{b,\mathrm{Cr}I} - E_{b,\mathrm{Ni}I} = 0.01$~eV for Ni--1Cr~\cite{hackett2008} and $E_{b,\mathrm{k}I} = E_{b,\mathrm{Ni}I}$ ($\beta_{kI}=1$) for Ni--1Fe and Ni--1Mn. Thus no preferential SIA-mediated transport contribution is introduced for Ni--1Fe and Ni--1Mn, which is justified by the repulsive or weak solute--dumbbell interactions~\cite{toijer_solute-point_2021}.
The resulting vacancy and SIA PDC ratios are shown in Fig.~\ref{fig:diff_data}b.

The TS energy for Mn was taken from the equilibrium GB segregation spectrum computed by Tuchinda et al.~\cite{tuchinda2025augmented} using the augmented potential method, which yields a peak segregation energy of $\Delta E^{\text{seg}}_{\text{Mn}} \approx -5$~kJ/mol for Ni--Mn in the dilute limit (Fig.~S4a).
This weak negative segregation energy produces a predicted equilibrium GB concentration of $\sim$3.1~at.\% at 500~K and $\sim$1.5~at.\% at 1473~K (Fig.~S4b), consistent with the experimentally observed pre-irradiation Mn enrichment.
No TS energies were applied for Fe or Cr ($\Delta E^{\text{seg}}_{\text{Fe}} = \Delta E^{\text{seg}}_{\text{Cr}} = 0$), consistent with the absence of measurable segregation in the unirradiated regions of those alloys (see Sec.~\ref{sec:results_exp}).

Unless otherwise stated, the bulk sink strength was taken to arise from a dislocation network with line density $\rho_b = 10^{14}$~m$^{-2}$ and point-defect capture efficiency $Z = 1$, giving $k_b^2 = \rho_b Z = 10^{14}$~m$^{-2}$.
For the 800~K simulations, an elevated effective bulk sink strength of $k_b^2 = 4 \times 10^{17}$~m$^{-2}$ was used to reproduce the experimentally observed suppression of RIS at this temperature. 
This value reflects the higher density of irradiation-generated microstructural sinks---including dislocation loops, voids, and sub-nanometer defect clusters---that develop in greatest density at intermediate temperatures in ion-irradiated Ni-based alloys~\cite{etienne2010,briggs2016,lu2017}, and which compete directly with GBs for point-defect flux and solute redistribution~\cite{jiao2011,dai2022,wang2022}. 
% The TEM-resolved loop and void populations reported for Ni--Cr alloys at comparable conditions correspond to sink strengths of order $10^{14}$--$10^{15}$~m$^{-2}$~\cite{barr2015,briggs2016}, three orders of magnitude below the $4\times10^{17}$~m$^{-2}$ value used here. We attribute this gap to sub-nanometer defect clusters below the TEM detection limit, which positron annihilation spectroscopy has shown can reach densities up to two orders of magnitude higher than TEM-visible populations at comparable dose~\cite{song2024dose}. While this is an indirect estimate, future work might utilize direct quantification of the sub-resolution cluster population and cluster dynamics modeling.

The depth-dependent damage rate for ion-irradiation simulations was obtained from SRIM calculations in Kinchin--Pease mode and fit to a Pearson IV function (Fig.~S5) used in the phase-field model.
% Reported doses correspond to the nominal SRIM/Kinchin--Pease (NRT-equivalent) damage, used directly as the point-defect source in the rate-theory model; the sensitivity of the predicted GB concentrations to the effective production efficiency of freely migrating defects is assessed in Supplementary Material Sec.~S8.
Reported doses correspond to the nominal SRIM/Kinchin--Pease damage, equivalent to the Norgett--Robinson--Torrens (NRT) standard~\cite{stoller2013use}; this nominal dose is used directly as the point-defect source in the rate-theory model, and the sensitivity of the predicted GB concentrations to the effective production efficiency of freely migrating defects is assessed in Supplementary Material Sec.~S8.
Separate 1D rate-theory simulations (see Fig.~S6 in Supplementary Material) were performed to confirm that vacancy and SIA concentrations at 5~$\mu$m depth are negligible relative to those near the damage peak, supporting the use of this depth as an unirradiated/thermal internal reference for TS.

All model parameters are summarized in Table~S1 in the Supplementary Material.

\subsection{Implementation}
\label{sec:implementation}

The rate-theory and phase-field model was implemented in the MOOSE framework using the numerical methods described in our previous work~\cite{kadambi2025}.
Time integration used the second-order backward differentiation formula (BDF2), and the nonlinear system was solved using a preconditioned Jacobian-free Newton--Krylov (PJFNK) method.
Linear Lagrange shape functions were used for the finite-element discretization, with a mesh-converged element size of 0.1~nm for a GB width of $\delta = 1$~nm.
Simulations employed a bicrystal domain with the GB at the center and periodic boundary conditions at the domain edges.
For ion-irradiation simulations, the depth-dependent damage rate was prescribed using the Pearson IV parameterization described in Sec.~\ref{sec:parameterization} and Fig.~S5.
For Ni--1Mn, a preliminary TS simulation at 1473~K for 48~h was first performed to reproduce the pre-irradiation heat-treatment history prior to subsequent irradiation simulations.
The initial condition for these simulations consisted of a uniform nominal solute concentration, $c_X(x,t=0)=c_X^{\mathrm{nominal}}$, and a hyperbolic-tangent order-parameter profile centered on the GB.
The resulting segregation profile was then used as the initial condition for subsequent TS, RETS, and RIS+RETS simulations at 500, 800, and 1000~K, as well as for the moving-GB calculations.
GB migration was studied in a 1D bicrystal by applying a free-energy driving force to one of the grains along with a GB mobility of $L_\eta \approx 10^{-13}\textup{--}10^{-11}$~m$^3$/(J\,s) (Sec.~\ref{subsec:model_results:moving_GB}).
2D bicrystal simulations were performed to verify that GB segregation in the unirradiated region at 5~$\mu$m depth is unaffected by irradiation effects at shallower depths.

\section{Results} \label{sec:results}

\subsection{Experimental} \label{sec:results_exp}
\begin{figure}[htp!] 
\centering
        \includegraphics[width=0.6\textwidth]{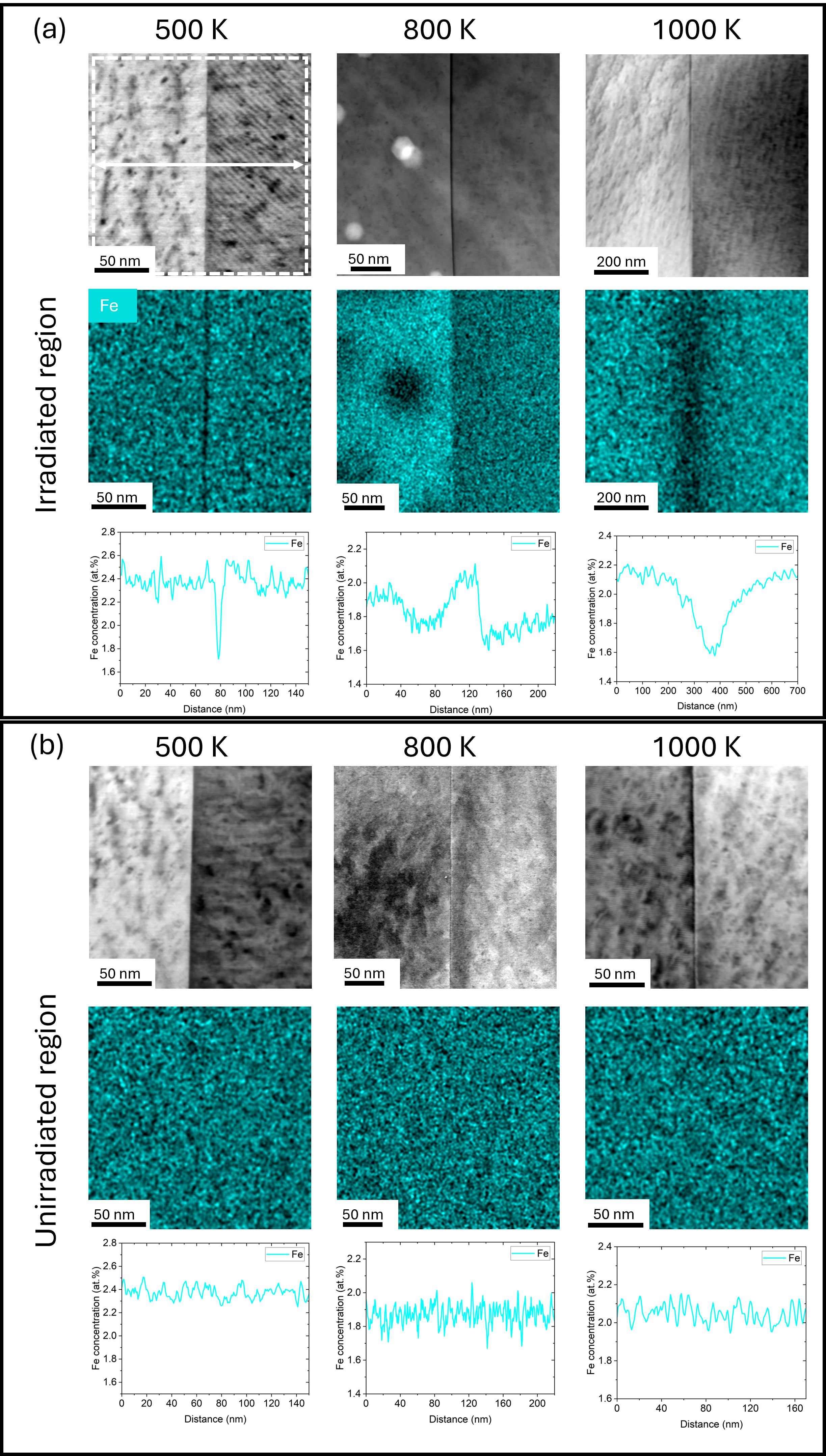}
    \caption{Compositional measurements at a GB in Ni--1Fe model alloys at the three temperatures investigated (500, 800, and 1000~K).
(a) STEM bright-field images, elemental maps, and Fe composition profiles recorded at high-angle GBs within the irradiated region ($\sim$1~$\mu$m).
(b) Corresponding measurements from the unirradiated region ($\sim$5~$\mu$m).
The line and integrated-width composition profiles were extracted as indicated by the white lines in (a).}
    \label{fig:exp:NiFe}
\end{figure}

\begin{figure}[htp!]
\centering
        \includegraphics[width=0.6\textwidth]{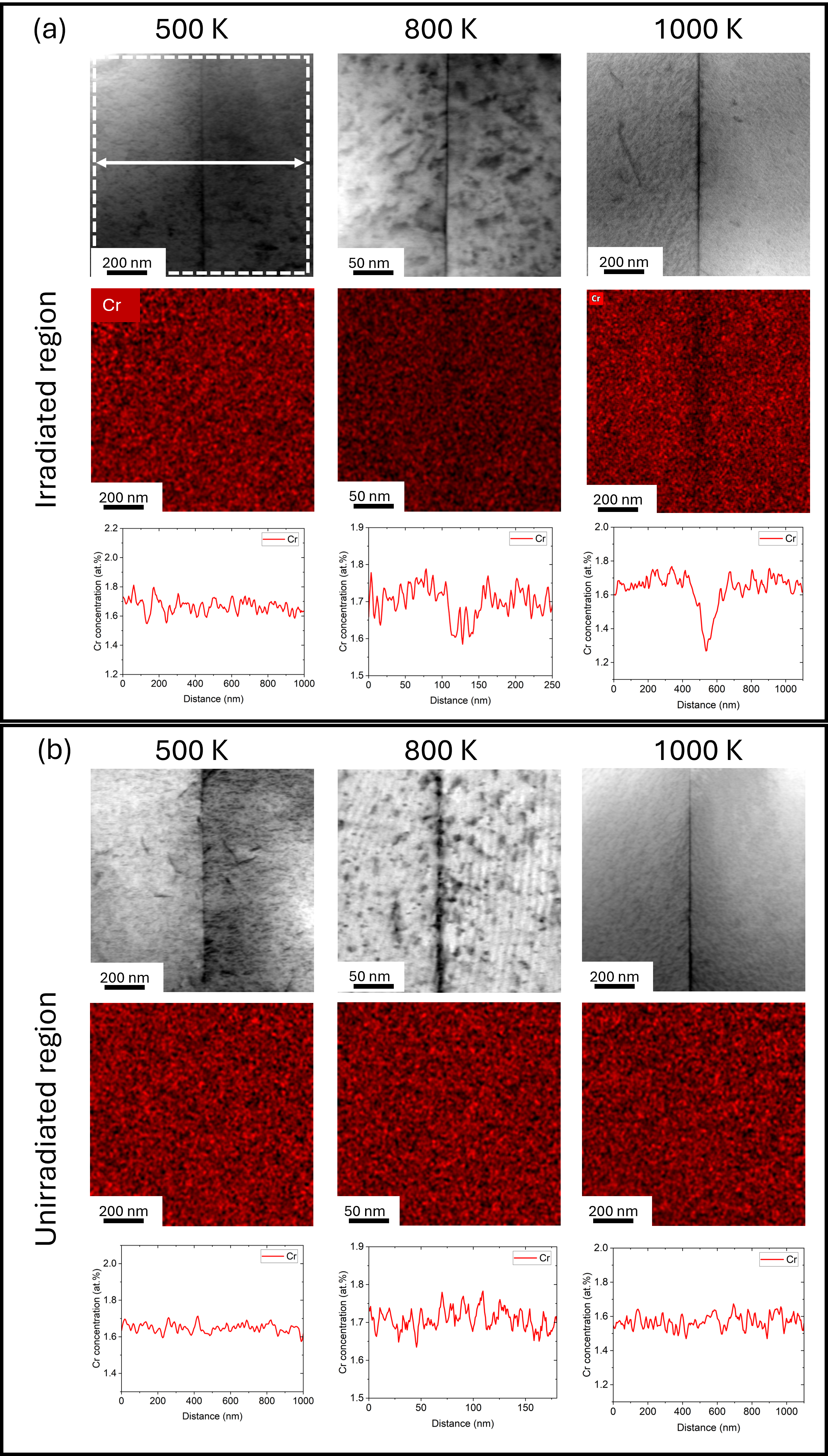}
    \caption{Compositional measurements at a GB in Ni--1Cr model alloys at the three temperatures investigated (500, 800, and 1000~K).
(a) STEM bright-field images, elemental maps, and Cr composition profiles recorded at high-angle GBs within the irradiated region ($\sim$1~$\mu$m).
(b) Corresponding measurements from the unirradiated region ($\sim$5~$\mu$m).
The line and integrated-width composition profiles were extracted as indicated by the white lines in (a).}
    \label{fig:exp:NiCr}
\end{figure}

Elemental analysis was performed using STEM-EDS to characterize chemistry in irradiated ($\sim$1~$\mu$m) and unirradiated regions ($\sim$5~$\mu$m) of the Ni--1Fe, Ni--1Cr, and Ni--1Mn model alloys.
In Ni--1Fe, Fe depletion is observed at GBs in the irradiated region at all three temperatures investigated (500, 800, and 1000~K), as shown in Fig.~\ref{fig:exp:NiFe}a.
At 500~K, the depletion is localized to a narrow region at the GB.
At 800~K, localized Fe-depleted regions are also associated with nearby voids in the irradiated region.
At 1000~K, the Fe depletion profile is broader and more diffuse, extending over several hundred nanometers.
In contrast, the unirradiated region at all three temperatures (Fig.~\ref{fig:exp:NiFe}b) shows no measurable Fe segregation at the GB.
In Ni--1Cr, the Cr concentration remains relatively uniform across the GB at 500~K in both irradiated and unirradiated regions (Fig.~\ref{fig:exp:NiCr}).
At 800 and 1000~K, the irradiated region (Fig.~\ref{fig:exp:NiCr}a) exhibits measurable Cr depletion at the GB plane.
In these cases, the Cr concentration decreases from a matrix value of approximately 1.65--1.75 at.\% to a minimum of approximately 1.30--1.60 at.\% at the GB core, with depletion widths on the order of 100--150~nm.
Unlike Ni--1Fe, no nearby voids are observed in the analyzed Ni--1Cr GB regions at 800~K.
The unirradiated regions (Fig.~\ref{fig:exp:NiCr}b) remain compositionally homogeneous at all temperatures.
Overall, the Ni--1Cr results indicate negligible segregation at 500~K and measurable Cr depletion at 800 and 1000~K in the irradiated region, with the strongest depletion at 1000~K.

\begin{figure}[htp!]
\centering
        \includegraphics[width=0.6\textwidth]{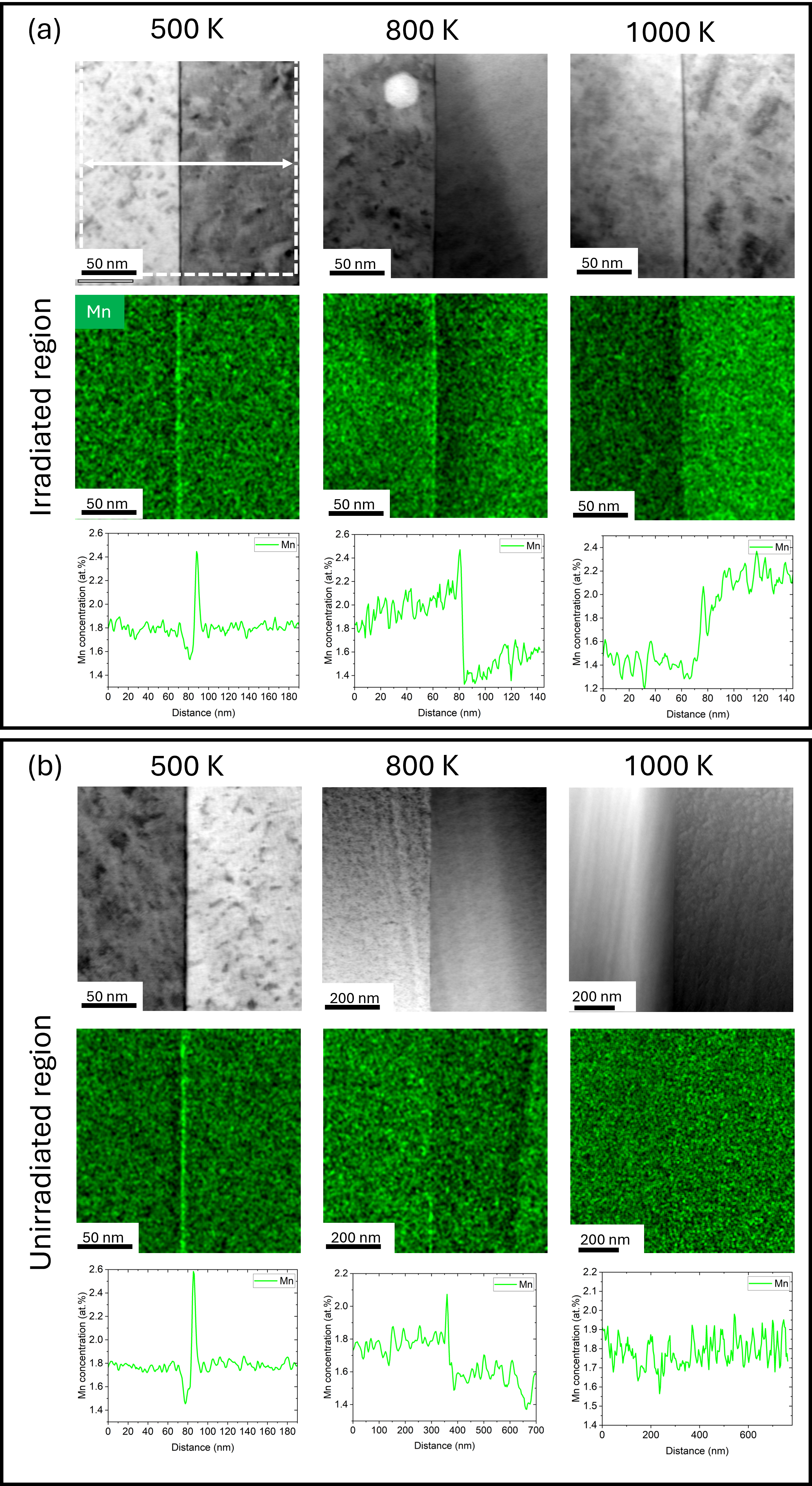}
    \caption{Compositional measurements at a GB in Ni--1Mn model alloys at the three temperatures investigated (500, 800, and 1000~K).
(a) STEM bright-field images, elemental maps, and Mn composition profiles recorded at high-angle GBs within the irradiated region ($\sim$1~$\mu$m).
(b) Corresponding measurements from the unirradiated region ($\sim$5~$\mu$m).
The line and integrated-width composition profiles were extracted as indicated by the white lines in (a).}
    \label{fig:exp:NiMn}
\end{figure}

In Ni--1Mn, the irradiated region at 500~K (Fig.~\ref{fig:exp:NiMn}a) shows a narrow Mn enrichment peak of approximately 2.4--2.6 at.\% superimposed on a matrix background of approximately 1.7--1.8 at.\%, with a small local minimum adjacent to the peak in several scans.
A similar enrichment profile is also observed in the unirradiated region at 500~K (Fig.~\ref{fig:exp:NiMn}b).
At 800~K, the Mn profile becomes markedly asymmetric.
In the irradiated region, Mn increases toward the GB to approximately 2.0--2.1 at.\%, reaches a sharp interfacial peak of approximately 2.4--2.5 at.\%, and then decreases abruptly on the opposite side of the GB to approximately 1.35--1.55 at.\%.
A similar but weaker asymmetric feature is also present in the unirradiated region at 800~K, where an Mn peak of approximately 2.0--2.1 at.\% coincides with the GB and is accompanied by a step-like decrease across the interface.
A nearby void is also observed in the irradiated region at 800~K, with corresponding local Mn depletion in the EDS profiles.
At 1000~K, the unirradiated region does not show a distinct Mn enrichment peak; instead, the profile exhibits a shallow GB-centered minimum extending over several hundred nanometers, decreasing to approximately 1.65--1.70 at.\% from a matrix level of approximately 1.8--1.9 at.\%.
In the irradiated region at 1000~K, the Mn concentration varies strongly between the two grains adjacent to the GB, with a step-like change from approximately 1.3--1.5 at.\% on one side to approximately 2.0--2.2 at.\% on the other, together with a slight interfacial maximum.
This profile asymmetry is consistent with GB displacement during irradiation.

\begin{figure}[htp!]
\centering
        \includegraphics[width=0.25\textwidth]{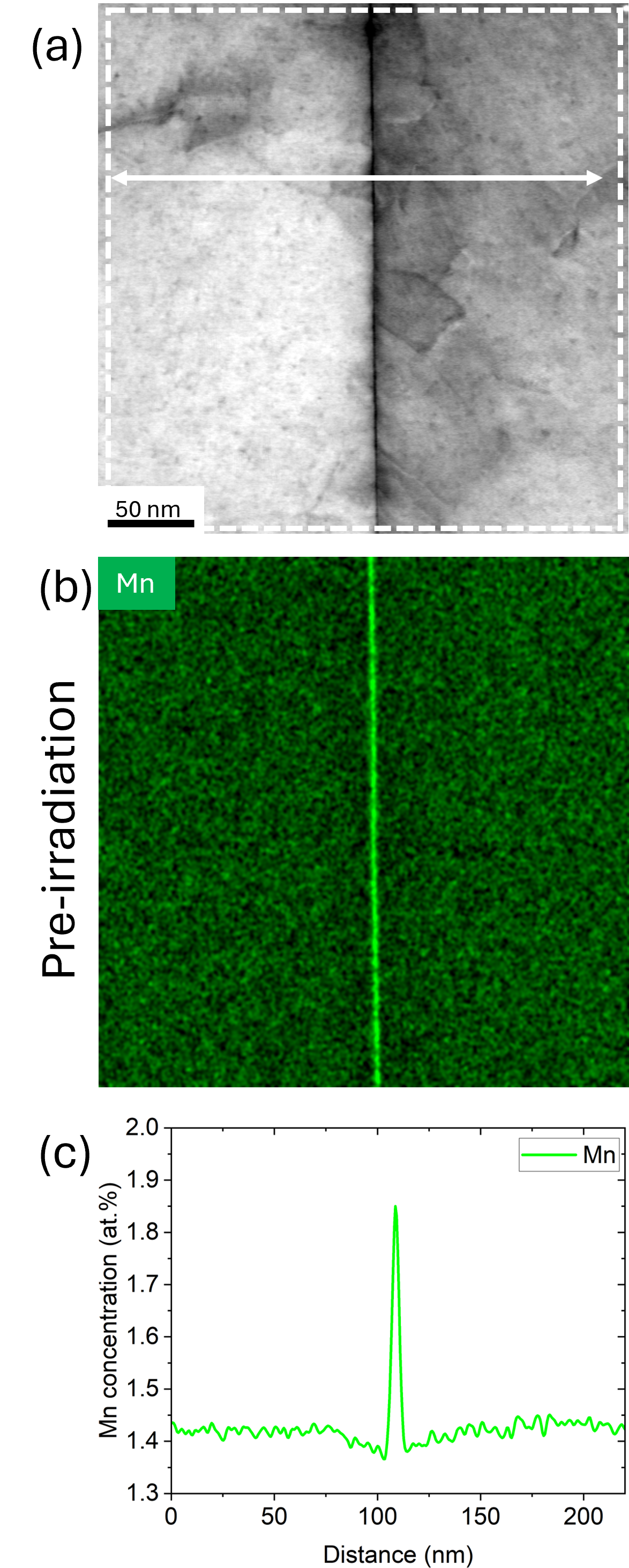}
    \caption{Compositional measurements at a GB in the same Ni--1Mn model alloy before irradiation.
(a) Bright-field image, (b) Mn elemental map, and (c) composition profiles recorded at a high-angle GB are shown.
The line and integrated-width composition profiles were extracted as indicated by the white lines in (a).}
    \label{fig:exp:NiMn_preirradiation}
\end{figure}

Fig.~\ref{fig:exp:NiMn_preirradiation} shows that Mn enrichment is already present at the GB in the Ni--1Mn alloy prior to irradiation.
This observation indicates that an initial segregation state exists before simultaneous irradiation and heating and must therefore be considered when interpreting the irradiated profiles.
The samples were heat treated at 1473 K for 48~h before water quenching, and this heat treatment is a likely origin of the pre-existing Mn enrichment.
The pre-irradiation enrichment profile is symmetric about the GB plane.
This pre-existing enrichment must be accounted for as an initial condition when interpreting irradiated profiles; notably, the profile asymmetry that develops at 800 and 1000~K after irradiation (Fig.~\ref{fig:exp:NiMn}a) is indicative of GB migration during the irradiation thermal hold.

Taken together, the three model alloys exhibit distinct and temperature-dependent GB segregation behaviors: Fe and Cr show irradiation-induced depletion, Mn shows pre-existing enrichment that evolves asymmetrically under irradiation, and all three alloys show evidence of voids at 800~K and depletion at the void surfaces.

\subsection{Modeling} \label{sec:results_mod}

\subsubsection{Simulation results for Ni-1Fe}
\label{subsec:model_results:NiFe}
\begin{figure}[htp!] 
\centering
    \begin{subfigure}[t]{0.48\textwidth}
        \includegraphics[width=1\textwidth]{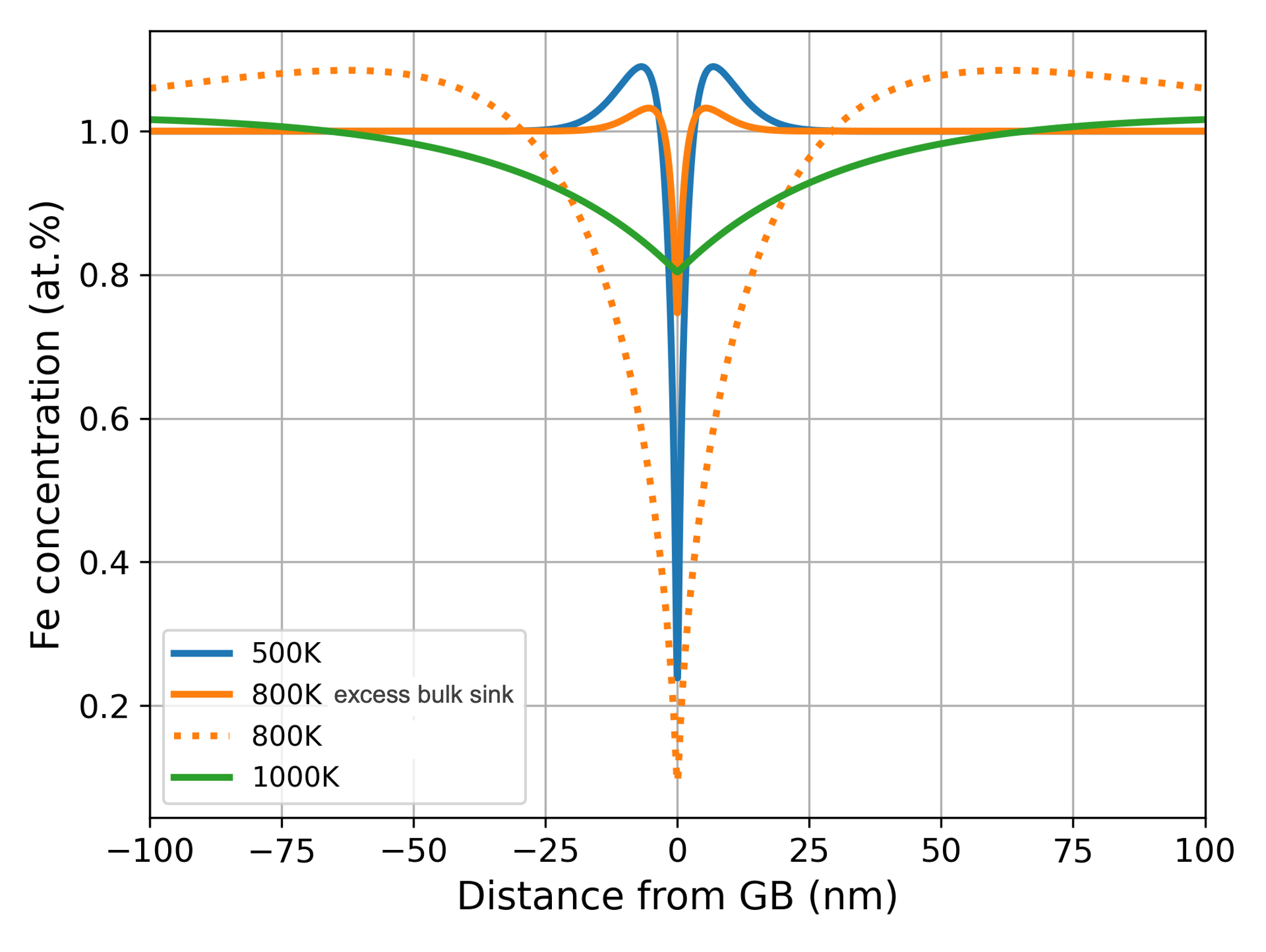}
    \caption{}
    \label{subfig:pf_NiFe_line_profile}
    \end{subfigure}
    \begin{subfigure}[t]{0.46\textwidth} 
        \includegraphics[width=1\textwidth]{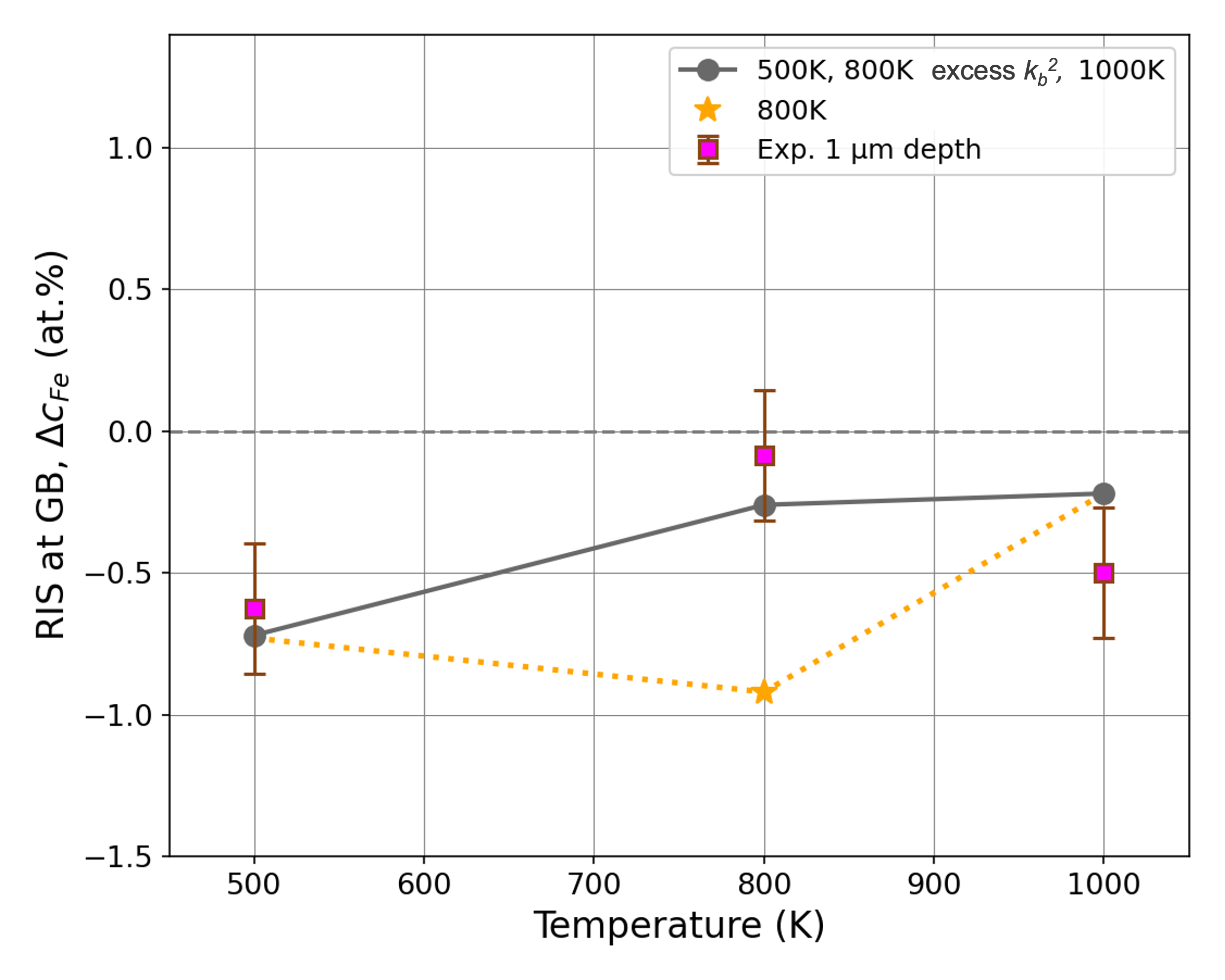}
    \caption{}
    \label{subfig:pf_NiFe_GB_delta}
    \end{subfigure}
    \caption{Model predictions of Fe RIS in Ni--1Fe at 2~dpa.
(a) Concentration profiles showing Fe depletion at GBs at 500~K, 800~K (with and without mean-field bulk sinks), and 1000~K.
(b) Temperature dependence of GB center concentration change ($\Delta c_{Fe}$); experimental values (squares) correspond to measurements at $\sim$1~$\mu$m depth.
Inclusion of bulk sinks reduces the predicted RIS at 800~K and improves agreement with experiment.}
    \label{fig:pf_NiFe}
\end{figure}

Fig.~\ref{fig:pf_NiFe} shows the model predictions of RIS in Ni--1Fe at 500, 800, and 1000~K.
The concentration profiles in Fig.~\ref{fig:pf_NiFe}a show Fe depletion at the GB at all temperatures.
At 500~K, the predicted depletion zone is moderate in magnitude and confined to a narrow region near the GB, in good agreement with the experimental profile (Fig.~\ref{fig:exp:NiFe}a).
At 800~K, the predicted depletion magnitude and width increase for the default bulk sink strength ($k^2_b=10^{14}$ m$^{-2}$), whereas inclusion of excess bulk sink strength ($k^2_b=4\times10^{17}$ m$^{-2}$, justified in Sec.~\ref{sec:bulk_sinks}) substantially reduces both.
At 1000~K, the profile becomes wider but shallower.
Beyond the depleted region, weak enrichment develops in the adjacent bulk as Fe redistributes away from the GB.
The steady-state point defect concentration profiles across the GB (Fig.~S7) show that the vacancy depletion zone width increases with temperature, directly influencing the width of the predicted segregation profiles.

The concentration change at the GB, defined as $\Delta c_{Fe} = c^{gb}_{Fe} - c^{bulk}_{Fe}$ (where $c^{gb}_{Fe}$ is the concentration at the GB center and $c^{bulk}_{Fe}\approx1$ at.\% is the far-field bulk concentration), is plotted as a function of temperature in Fig.~\ref{fig:pf_NiFe}b for direct comparison with experimental values. $\Delta c_{Fe}$ is negative at all temperatures, consistent with Fe depletion ($\Delta E^{\text{seg}}_{\text{Fe}} = 0$, so no TS driving force modifies RIS).
With the default bulk sink strength, predictions show a non-monotonic temperature dependence, with the greatest depletion at 800~K and the least at 1000~K.
% Including the excess bulk sink strength at 800~K (reflecting the higher density of voids and defect clusters that develop at intermediate temperatures) substantially reduces the predicted depletion and improves agreement with the experimental measurements in Fig.~\ref{fig:exp:NiFe}.
Including the excess bulk sink strength at 800~K substantially reduces the predicted depletion and improves agreement with the experimental measurements in Fig.~\ref{fig:exp:NiFe}.
% The predicted $\Delta c_{Fe}$ values are in reasonable agreement with experimental measurements across all three temperatures, with the model with enhanced 800~K bulk sinks improving agreement at 800~K.

\subsubsection{Simulation results for Ni-1Cr}
\label{subsec:model_results:NiCr}

\begin{figure}[htp!] 
\centering
    \begin{subfigure}[t]{0.48\textwidth}
        \includegraphics[width=1\textwidth]{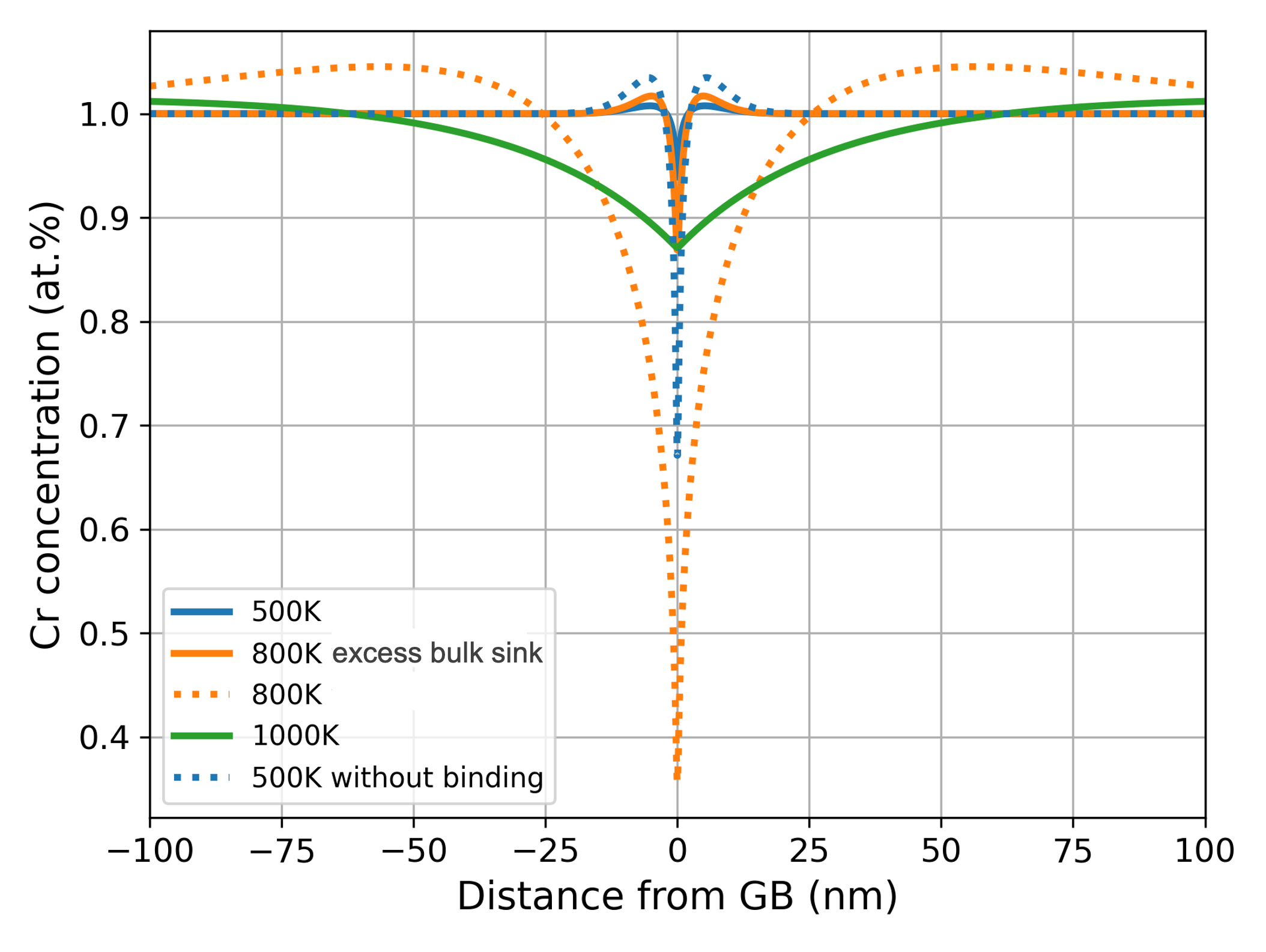}
    \caption{}
    \label{subfig:pf_NiCr_line_profile}
    \end{subfigure}
    \begin{subfigure}[t]{0.46\textwidth} 
        \includegraphics[width=1\textwidth]{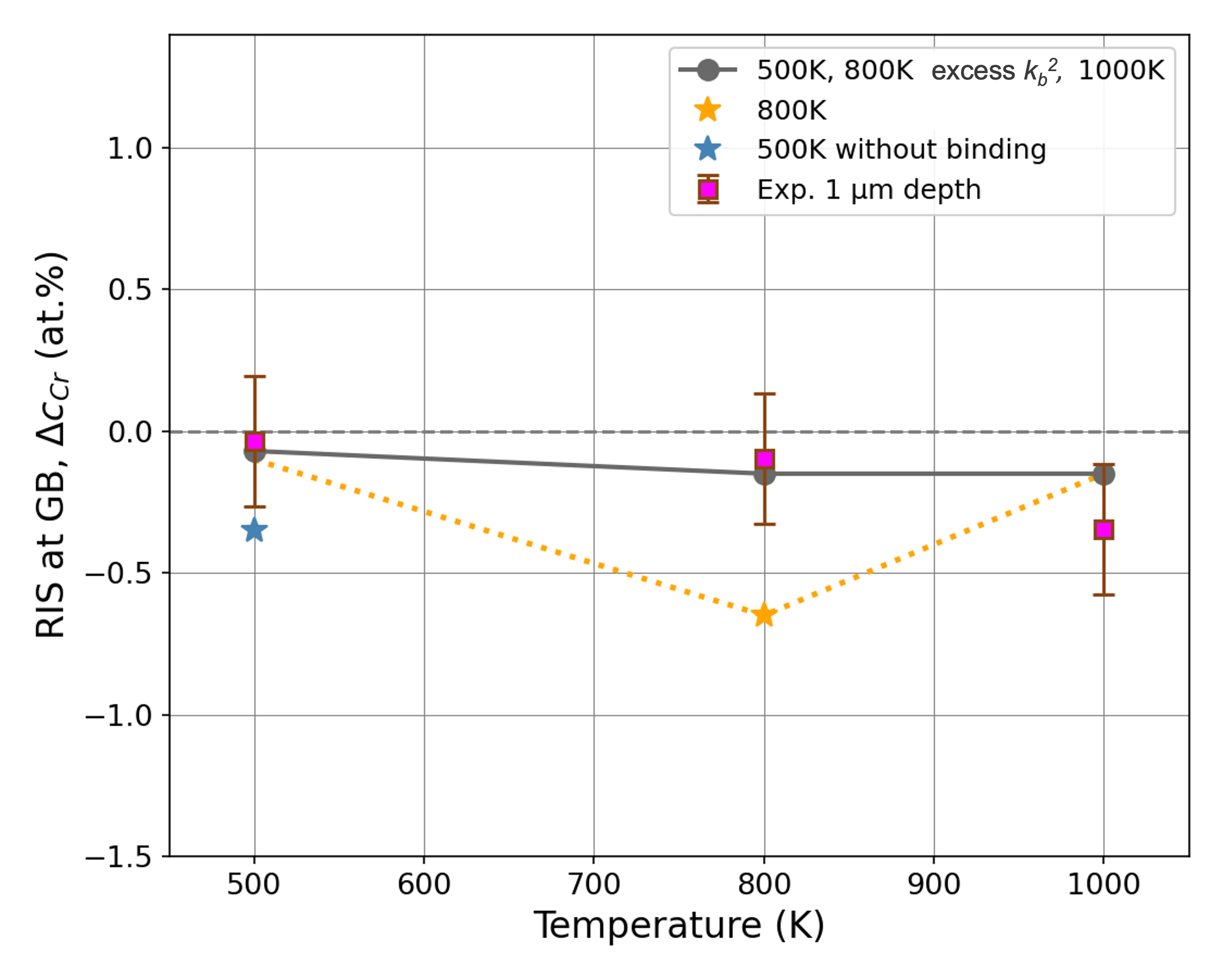}
    \caption{}
    \label{subfig:pf_NiCr_GB_delta}
    \end{subfigure}
    \caption{Model predictions of Cr RIS in Ni--1Cr at 2~dpa.
(a) Concentration profiles showing Cr depletion at GBs at 500~K (with and without Cr--SIA binding), 800~K (with and without mean-field bulk sinks), and 1000~K.
(b) Temperature dependence of GB center concentration change ($\Delta c_{Cr}$); experimental values (squares) correspond to measurements at $\sim$1~$\mu$m depth.
SIA-mediated transport suppresses the low-temperature depletion, while bulk sinks reduce the predicted RIS at 800~K.}
    \label{fig:pf_NiCr}
\end{figure}

Fig.~\ref{fig:pf_NiCr} shows model predictions of RIS in Ni--1Cr at 2 dpa.
The concentration profiles in Fig.~\ref{fig:pf_NiCr}a predict Cr depletion at all temperatures.
When only preferential vacancy-mediated transport is considered (as done for Fe with $E_{b,\mathrm{Fe}I} = E_{b,\mathrm{Ni}I}$, describing no preference in SIA-mediated transport between the solute and Ni), Cr depletion at 500~K is smaller than the corresponding Fe depletion because the vacancy PDC ratio for Cr is lower than that for Fe (Fig.~\ref{fig:diff_data}).
When preferential Cr transport by the SIA mechanism is included (using a relative Cr--SIA binding energy of $E_{b,\mathrm{Cr}I} - E_{b,\mathrm{Ni}I} = 0.01$~eV), the predicted depletion at 500~K becomes nearly negligible, in agreement with the experimental observation in Fig.~\ref{fig:exp:NiCr}.
Preferential Cr--SIA binding drives Cr toward the GB via interstitial flux, partially offsetting vacancy-driven depletion and resulting in near-zero net segregation at 500~K.
At 800 and 1000~K, the Cr--SIA binding contribution becomes negligible and the profiles are dominated by vacancy-mediated depletion.
As in Ni--1Fe, excess bulk sink strength at 800~K substantially reduces the predicted depletion magnitude and width.
At 1000~K, the profile is broader and shallower than at 800~K.

The GB center concentration change, defined as $\Delta c_{Cr} = c^{gb}_{Cr} - c^{bulk}_{Cr}$, is shown in Fig.~\ref{fig:pf_NiCr}b. $\Delta c_{Cr}$ is nearly negligible at 500~K and becomes more strongly negative at 800 and 1000~K, indicating near-zero segregation at low temperature and clear Cr depletion at higher temperatures.
The magnitude of depletion is smaller than that predicted for Fe at all temperatures, consistent with the lower vacancy-mediated Cr transport and the partial cancellation by SIA-mediated flux at low temperature.
The predicted $\Delta c_{Cr}$ values are in reasonable agreement with experimental measurements across all three temperatures: inclusion of Cr--SIA binding brings the 500~K prediction into agreement with the near-zero experimental segregation, and the model with enhanced 800~K bulk sinks brings predictions into closer agreement with experiment at 800~K.

\subsubsection{Simulation results for Ni-1Mn} \label{subsec:model_results:NiMn_stat}

Unlike Fe and Cr, for which RIS is the sole irradiation-driven segregation mechanism, Mn possesses a thermodynamic driving force for GB enrichment as seen from the pre-irradiated experimental characterization result in Fig.~\ref{fig:exp:NiMn_preirradiation}.
To isolate the individual mechanistic contributions, three types of simulations were performed: (1)~RIS-only, in which only the preferential vacancy transport of Mn is active ($\Delta E^{\text{seg}}_{\text{Mn}} = 0$ for TS and $E_{b,\mathrm{Mn}I} = E_{b,\mathrm{Ni}I}$ for SIA transport); (2)~RETS-only, in which the segregation energy ($\Delta E^{\text{seg}}_{\text{Mn}} \approx -5$~kJ/mol) is active and irradiation accelerates diffusion but no preferential vacancy-mediated transport is active (i.e. the vacancy partial diffusivities for Mn and Ni are set to be equal as $d_{MnV} = d_{NiV}$); and (3)~combined RIS+RETS, in which both mechanisms operate simultaneously under irradiation-enhanced kinetics.
A pre-irradiation TS simulation at 1473~K for 48~h was first performed to reproduce the experimental heat-treatment history, and the resulting profile served as the initial condition for the RETS-only and RIS+RETS simulations of unirradiated (5~$\mu$m) and irradiated (1.5~$\mu$m) regions; the RIS-only case was initialized from a uniform nominal Mn concentration, consistent with its zero segregation energy.
Using the segregation energy from Tuchinda et al.~\cite{tuchinda2025augmented}, the equilibrium GB Mn concentration in Ni--Mn decreases from approximately 3.1~at.\% at 500~K to 1.5~at.\% at 1473~K (Fig.~S4b), which provides a thermal baseline for interpreting segregation under irradiation.
 
\begin{figure}[htp!] 
\centering
    \begin{subfigure}[t]{0.48\textwidth}
        \includegraphics[width=1\textwidth]{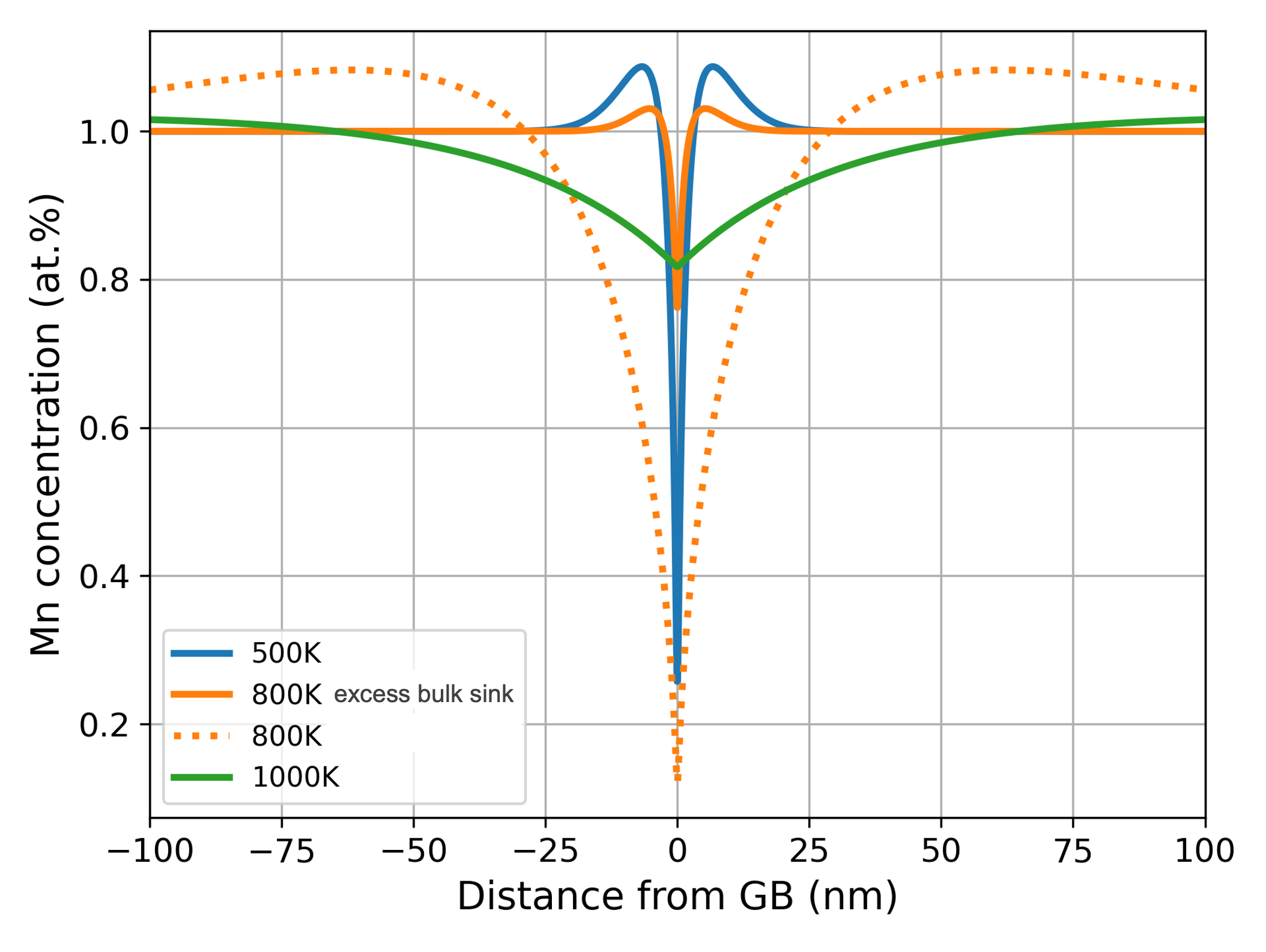}
    \caption{RIS}
    \label{subfig:pf_NiMn_RIS_line_profile}
    \end{subfigure}
    \begin{subfigure}[t]{0.48\textwidth}
        \includegraphics[width=1\textwidth]{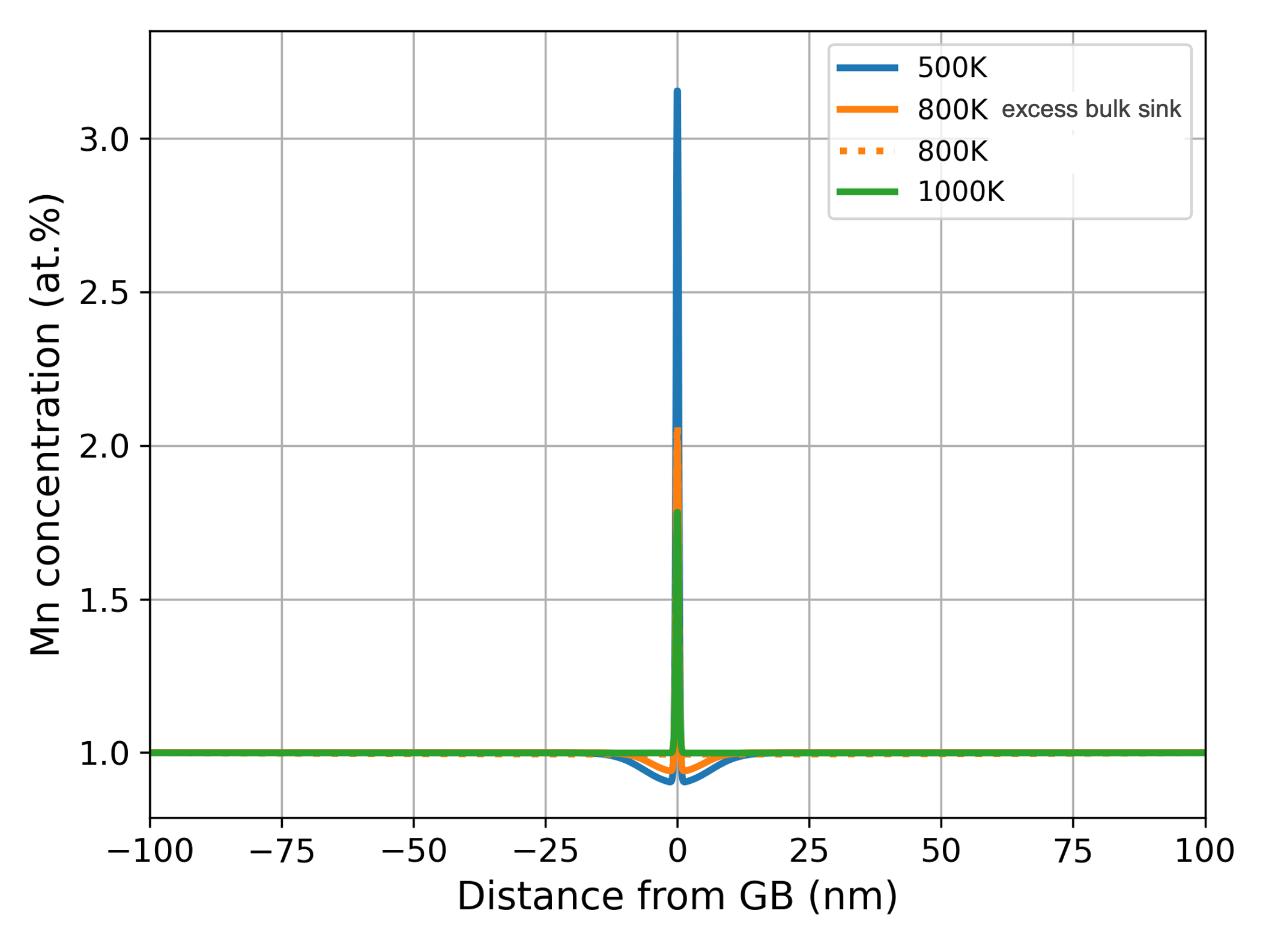}
    \caption{RETS}
    \label{subfig:pf_NiMn_RETS_line_profile}
    \end{subfigure}
    ~~
    \begin{subfigure}[t]{0.48\textwidth}
        \includegraphics[width=1\textwidth]{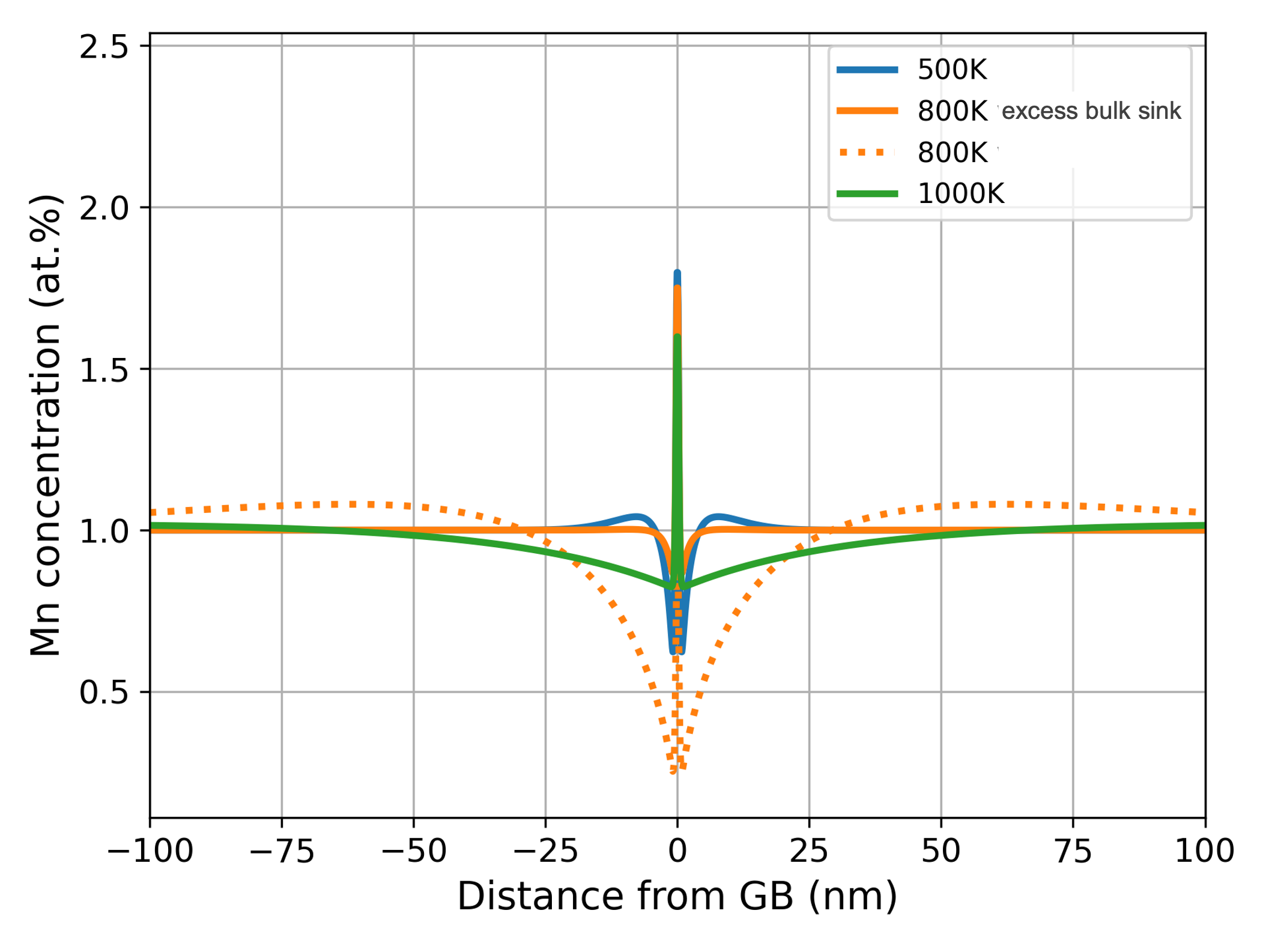}
    \caption{RIS+RETS}
    \label{subfig:pf_NiMn_RISandRETS_line_profile}
    \end{subfigure}
    \begin{subfigure}[t]{0.46\textwidth} 
        \includegraphics[width=1\textwidth]{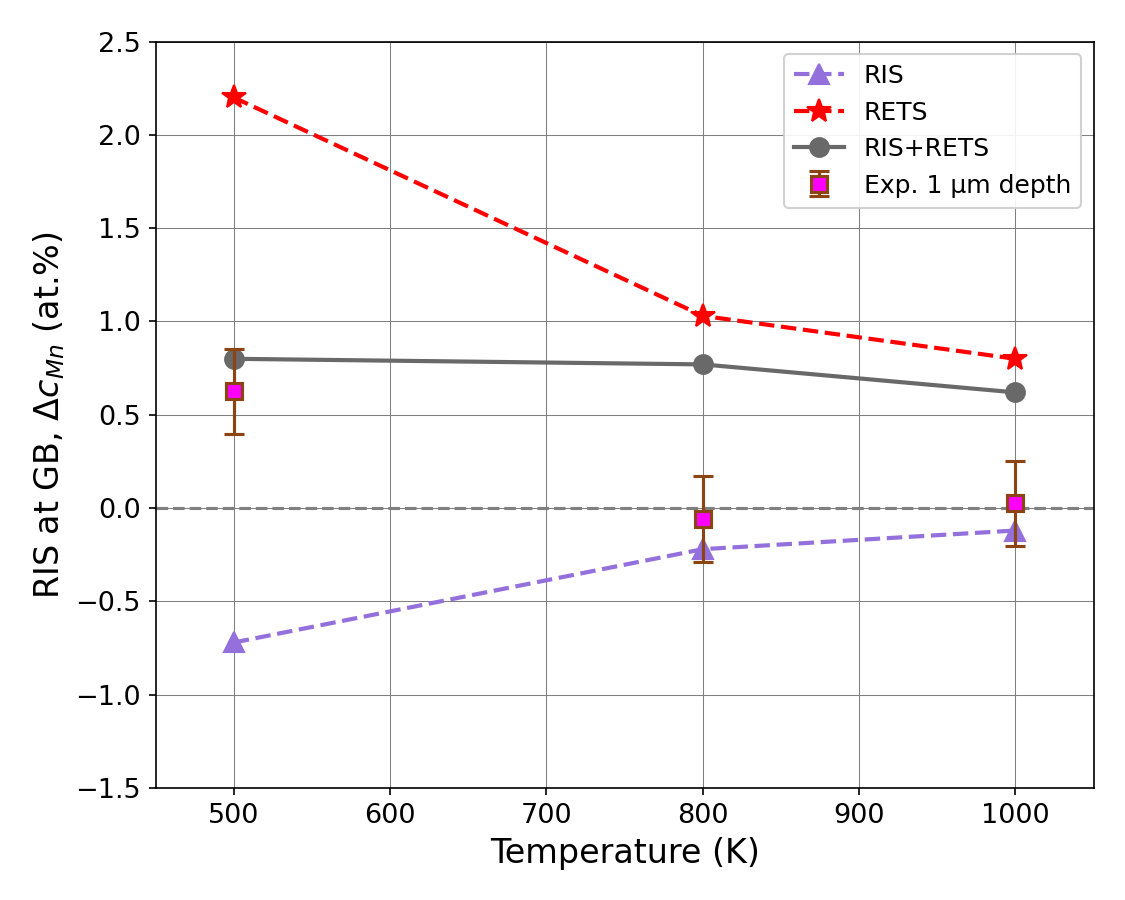}
    \caption{Concentration change at GB}
    \label{subfig:pf_NiMn_GB_delta}
    \end{subfigure}
    \caption{Model predictions of Mn segregation in Ni--1Mn at 500, 800, and 1000~K to 2~dpa.
(a) RIS-only profiles showing depletion.
(b) RETS-only profiles showing enrichment.
(c) Combined RIS+RETS profiles showing competing depletion and enrichment contributions. At 800~K, results with and without mean-field bulk sinks are shown.
(d) Temperature dependence of GB center concentration change ($\Delta c_{Mn}$): RIS predicts depletion, whereas RETS and RIS+RETS predict net enrichment. Experimental values (squares) correspond to measurements at $\sim$1~$\mu$m depth.}
    \label{fig:pf_NiMn}
\end{figure}

Fig.~\ref{fig:pf_NiMn} shows model predictions of RIS, RETS, and combined RIS+RETS in Ni--1Mn at 2 dpa.
The RIS-only results (Fig.~\ref{fig:pf_NiMn}a,d) predict Mn depletion at all temperatures, consistent with preferential vacancy-mediated Mn transport relative to Ni.
The RETS-only results (Fig.~\ref{fig:pf_NiMn}b,d) predict GB enrichment at all temperatures, with the largest enrichment at 500~K and the smallest at 1000~K.
% The RETS-only predictions are in reasonable agreement with the experimentally observed Mn enrichment profiles at 500 and 800~K in the unirradiated region, supporting the thermodynamic origin of the pre-irradiation segregation.
Compared with the RIS profiles, the RETS profiles are confined to a much narrower region near the GB and are only weakly affected by excess bulk sinks at 800~K.
% At 500~K, the RETS-only GB concentration exceeds the 1473~K pre-irradiation value because the equilibrium segregation itself is greater at lower temperature, reflecting the entropy-driven temperature dependence of $\Delta E^{\text{seg}}_{\text{Mn}}$ (Supplementary Material, Sec.~S4); RETS therefore drives further enrichment beyond the pre-existing 1473~K profile as the system evolves toward this lower-temperature equilibrium.

The combined RIS+RETS predictions in the irradiated 1.5~$\mu$m region in Fig.~\ref{fig:pf_NiMn}c produce non-monotonic profiles in which enrichment at the GB center is retained but reduced by the competing RIS contribution.
At 800~K, inclusion of excess bulk sink strength substantially reduces the combined GB enrichment, with RIS-driven depletion becoming less pronounced.
The predicted GB center concentration change, $\Delta c_{Mn} = c^{gb}_{Mn} - c^{bulk}_{Mn}$, remains positive for RETS-only and RIS+RETS cases, whereas the RIS-only case remains negative.
The RIS-only Mn profiles are qualitatively similar to the Fe depletion profiles in Ni--1Fe.
At 500~K, the RIS+RETS prediction is consistent with the experimental measurement in the irradiated region (Fig.~\ref{fig:exp:NiMn}a). 
% although the predicted depletion regions surrounding the enrichment peak are broader than observed experimentally.
At 800~K and 1000~K, the experimentally observed segregation profiles are asymmetric, whereas the stationary-GB simulations presented so far exhibit symmetric segregation; this discrepancy is examined through the moving-GB simulations explored in Sec.~\ref{subsec:model_results:moving_GB}.

\subsubsection{Depth-dependent segregation in ion-irradiated Ni--1Mn}
\label{subsec:model_results:NiMn_2D}

\begin{figure}[htp!]
    \centering
    \includegraphics[width=0.8\textwidth]{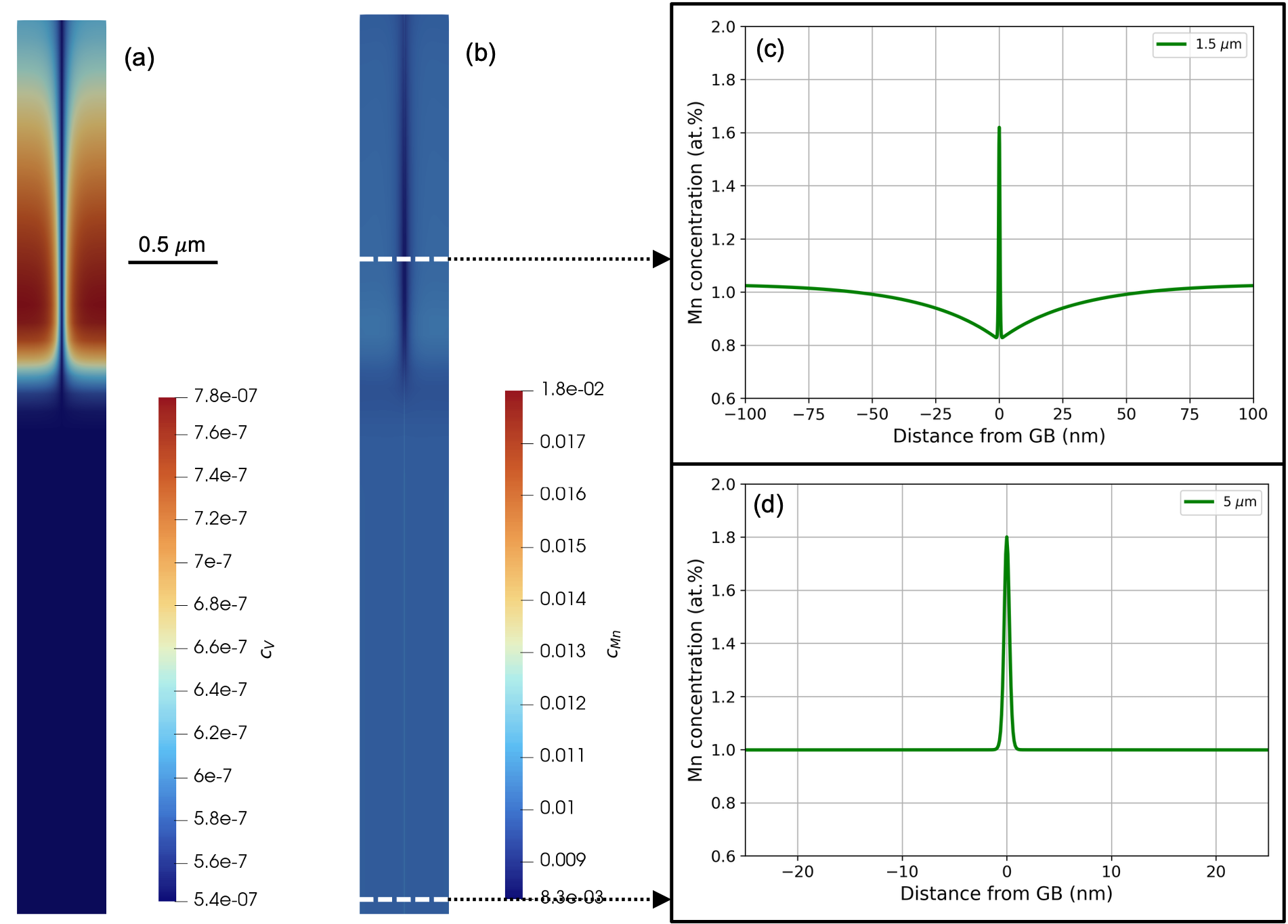}
    \caption{2D rate-theory phase-field simulation of a bicrystal Ni--1Mn.
(a) Vacancy concentration and (b) Mn concentration as a function of depth from the surface after ion irradiation at 1000~K for 0.37~h.
(c) Mn concentration profile across the GB at 1.5~$\mu$m depth (irradiated region).
(d) Mn concentration profile at 5~$\mu$m depth (unirradiated region).}
    \label{fig:pf_NiMn_2D}
\end{figure}

% To confirm the depth-dependent irradiation damage and that the resulting GB segregation in the irradiated region does not affect GB segregation in the 5~$\mu$m region, 2D bicrystal phase-field simulations of ion-irradiated Ni--1Mn were performed.
To confirm that the depth-dependent irradiation damage does not affect GB segregation in the 5~$\mu$m region, 2D bicrystal phase-field simulations of ion-irradiated Ni--1Mn were performed.
Fig.~\ref{fig:pf_NiMn_2D} shows the vacancy and Mn concentration fields as a function of depth from the irradiated surface after 0.37~h at 1000~K.
At $\sim$1.5~$\mu$m depth, corresponding to the damage peak, the Mn concentration profile across the GB (Fig.~\ref{fig:pf_NiMn_2D}c) exhibits combined RIS+RETS behavior with central enrichment superimposed on broader depletion regions, consistent with the 1D simulation profiles in Fig.~\ref{fig:pf_NiMn}.
% At 5~$\mu$m depth, in the unirradiated thermal region, the Mn profile (Fig.~\ref{fig:pf_NiMn_2D}d) reflects TS that is effectively unperturbed by irradiation-induced point-defect fluxes, consistent with the TS-only behavior captured by the concurrent moving-GB calculations in Sec.~\ref{subsec:model_results:moving_GB}.
At 5~$\mu$m depth, in the unirradiated thermal region, the Mn profile (Fig.~\ref{fig:pf_NiMn_2D}d) reflects TS that is effectively unperturbed by irradiation-induced point-defect fluxes.
% These results support the use of the 5~$\mu$m region as an internal RETS reference.
Supplementary simulations including enhanced GB diffusion (2--3 orders of magnitude faster than bulk diffusion) produced no measurable change in the segregation profile at 5~$\mu$m.
The defect damage and its impact on segregation thus do not extend beyond $\sim$2~$\mu$m depth, confirming that the 5~$\mu$m region serves as an unperturbed internal reference for TS.

\subsubsection{Segregation at moving GBs}
\label{subsec:model_results:moving_GB}

The stationary-GB simulations presented above produce symmetric segregation profiles, whereas the experimental measurements in Ni--1Mn show pronounced asymmetry at 800 and 1000~K.

To examine the origin of the profile asymmetry, concurrent moving-GB calculations were initialized from the post-1473~K thermodynamic-segregation state, after which RIS, RETS, and GB motion were evolved concurrently from $t=0$ to 1332~s (2 dpa). Equal bulk sink strengths were assigned to the two grains, and GB motion was driven by applying the $\Delta f$ term in Eq.~\ref{eq:eta_evol} to the right grain.

\begin{figure}[htp!]
    \centering
    \includegraphics[width=0.9\textwidth]{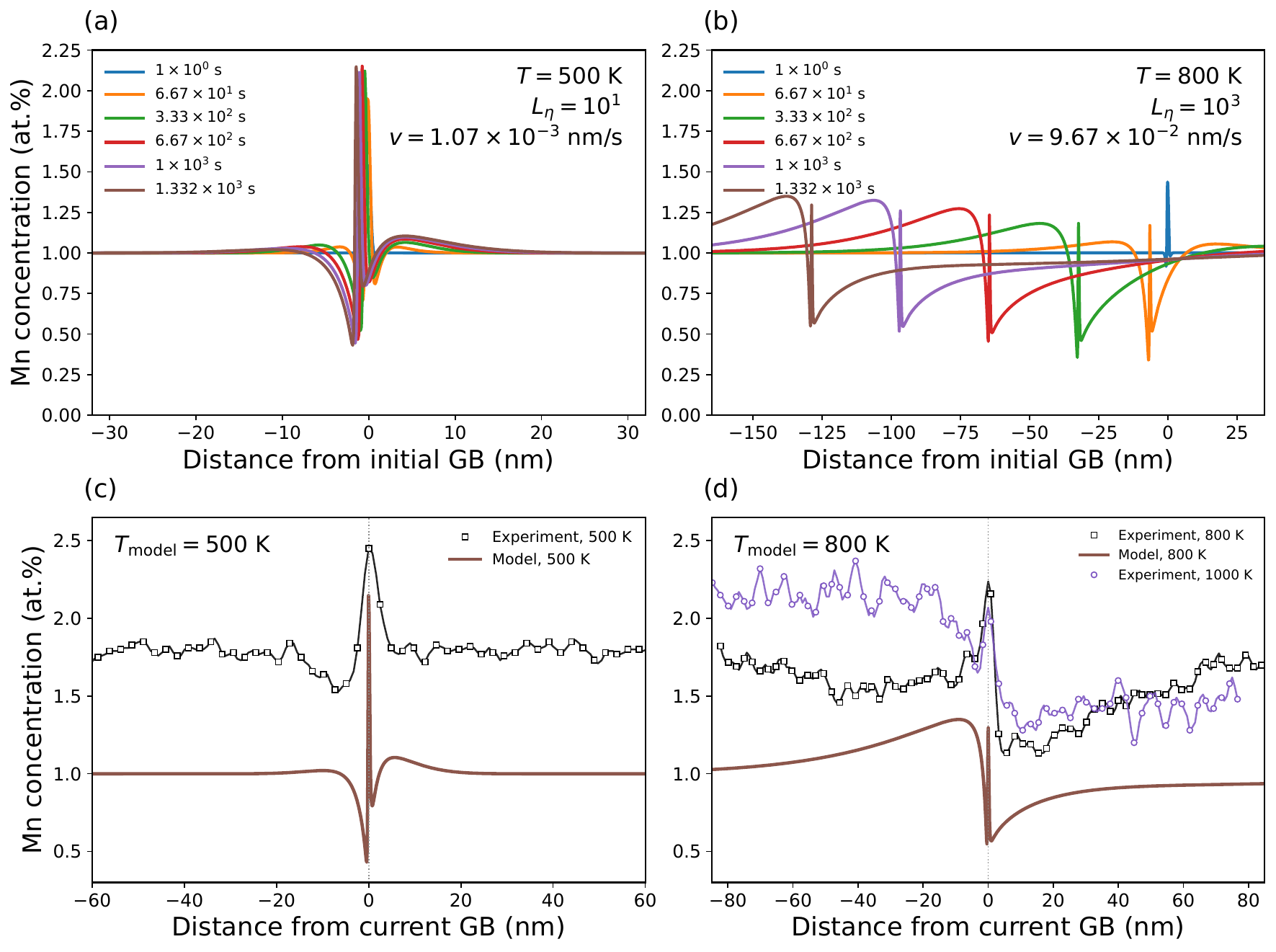}
    \caption{Concurrent RIS+RETS and GB-migration calculations for Ni--1Mn irradiated at 500 and 800~K following the 1473~K heat treatment. (a,b) Evolution of the calculated Mn concentration profiles at 500 and 800~K, respectively. (c,d) Calculated profiles at 1332~s (approximately 2 nominal dpa), centered on the current GB position and compared directly with the experimental profiles. The 1000~K experimental profile is also included in panel (d) to demonstrate that the asymmetric profile observed at 800~K persists at 1000~K.}
    \label{fig:concurrent_postHT_movingGB}
\end{figure}

Fig.~\ref{fig:concurrent_postHT_movingGB}a and b show the concurrent RIS+RETS evolution at 500 and 800~K, respectively. At 500~K, a GB mobility of $L_\eta\approx10^{-13}$~m$^3$/(J\,s) yields a GB velocity of approximately $10^{-3}$~nm/s. The resulting profile is weakly asymmetric, with a modest Mn-enriched region further ahead of the moving GB, a narrow depletion dip immediately ahead of the GB, and a sharp enrichment peak at the GB center. At 800~K, the higher GB mobility of $L_\eta\approx10^{-11}$~m$^3$/(J\,s) yields a GB velocity of approximately $10^{-1}$~nm/s, producing pronounced segregation asymmetry. A broad Mn-enriched region develops in the grain ahead of the moving GB, consistent with the RIS-only moving-GB signature, narrowing into a sharp depletion dip immediately ahead of the GB before the enrichment peak at the GB center; a depletion wake forms immediately behind the GB before gradually recovering toward the bulk concentration.
The absolute simulated concentration profiles at 1332~s are compared directly with the experimental profiles in Fig.~\ref{fig:concurrent_postHT_movingGB}c,d. The simulated profiles closely reproduce the principal experimental features at 500 and 800~K, including GB enrichment and the asymmetric enrichment/depletion fields. The 1000~K experimental profile is also included in Fig.~\ref{fig:concurrent_postHT_movingGB}d to demonstrate that the asymmetric profile observed at 800~K persists at 1000~K.

\section{Discussion} \label{sec:discussion}

The experimental observations, together with the modeling and simulation results presented in Sec.~\ref{sec:results}, reveal element-specific segregation behavior that cannot be explained by a single mechanism.
Collectively, these results help elucidate how the relative roles of vacancy-mediated transport (IK effect), SIA-mediated transport, TS, bulk sink competition, and GB migration jointly determine GB chemistry under irradiation.
In the following subsections, we discuss the mechanistic origins of the observed segregation for each solute (Sec.~\ref{sec:disc:NiFe}--\ref{sec:disc:NiMn}), the role of bulk sinks (Sec.~\ref{sec:bulk_sinks}), the effect of GB migration (Sec.~\ref{sec:gb_migration}), and the limitations of the present approach (Sec.~\ref{sec:limitations}).

\subsection{Vacancy-mediated RIS in Ni--Fe}
\label{sec:disc:NiFe}

Experimentally, irradiation produces consistent Fe depletion at GBs across all temperatures investigated (500, 800, and 1000~K) in Ni--1Fe (Fig.~\ref{fig:exp:NiFe}a).
No detectable composition change is observed in the unirradiated region at 5~$\mu$m depth (Fig.~\ref{fig:exp:NiFe}b), indicating that the Fe redistribution is driven primarily by RIS.
This behavior is consistent with the vacancy--atom exchange mechanism, or IK effect, in which preferential vacancy--Fe exchange drives Fe away from the GB sink.
The vacancy PDC ratio for Fe exceeds unity across the full temperature range investigated ($d_{\text{FeV}}/d_{\text{NiV}} \approx 2.7$ at 500~K, Fig.~\ref{fig:diff_data}b), indicating that Fe is always the faster-diffusing species via the vacancy mechanism \cite{toijer_solute-point_2021}.
The repulsive interaction between Fe and the $\langle 100 \rangle$ mixed Ni--Fe dumbbell is assumed to render SIA-mediated transport negligible \cite{tucker2010,toijer_solute-point_2021}, so the IK mechanism operates without a competing SIA contribution.
Rate-theory model predictions are in good agreement with the experimental observations, reproducing Fe depletion at all temperatures (Figs.~\ref{fig:pf_NiFe}a,b).
We note that the present measurements are performed at random high-angle GBs, and the depletion magnitude is expected to vary with GB character; coherent twin boundaries, for example, have been shown to exhibit significantly reduced RIS relative to random boundaries \cite{barr2015}, which introduces scatter in the experimental data that is not captured by the present work.
The Fe depletion observed around voids at 800~K (Fig.~\ref{fig:exp:NiFe}a) shows RIS occurring at multiple competing sinks (GBs and voids) simultaneously, and is discussed further in Sec.~\ref{sec:bulk_sinks}.

Comparison with prior studies in concentrated Ni-based alloys further supports the interpretation on Fe depletion under RIS.
Allen and Was~\cite{allen1998acta} reported Fe depletion at GBs in austenitic Fe--Cr--Ni alloys under proton irradiation and attributed the behavior to the IK mechanism.
The present Ni--1Fe results isolate the intrinsic Fe segregation tendency in fcc Ni without the confounding influence of Cr or other solutes, and thus provide experimental validation of the SCMF-pa\-ram\-e\-ter\-ized vacancy transport coefficients of Toijer et al.~\cite{toijer_solute-point_2021}.
Object kinetic Monte Carlo simulations by Barashev et al.~\cite{barashev2019chemically} have indeed shown that chemically biased vacancy--Fe exchange drives Fe depletion at sinks in Ni--Fe alloys. 
%consistent with the present experimental observations in the dilute limit.
% To the best of our knowledge, this is the first systematic experimental study of RIS in a dilute Ni--Fe binary alloy.

Predictions of the temperature dependence of Fe depletion (Fig.~\ref{subfig:pf_NiFe_GB_delta} with default bulk sink strength) follows a bell-shaped behavior \cite{was2007fundamentals,ardell2016}, with segregation suppressed below 500~K by limited defect mobility and diminished above 800~K by enhanced vacancy--SIA recombination and thermally activated back-diffusion.
The broader but shallower profiles predicted at 1000~K are also consistent with increased diffusion length scales.
The model reproduces the low-temperature narrow depletion and the high-temperature broad shallow profile.

\subsection{Vacancy- and SIA-mediated RIS in Ni--Cr}
\label{sec:disc:NiCr}

Experimentally, Ni--1Cr exhibits measurable Cr depletion at 800 and 1000~K (Fig.~\ref{fig:exp:NiCr}a), but no detectable segregation at 500~K in either the irradiated or unirradiated regions (Fig.~\ref{fig:exp:NiCr}).
The smaller depletion magnitude relative to Fe at equivalent temperatures is consistent with the lower vacancy PDC ratio for Cr ($d_{\text{CrV}}/d_{\text{NiV}} < d_{\text{FeV}}/d_{\text{NiV}}$ in Fig.~\ref{fig:diff_data}b) \cite{toijer_solute-point_2021,tucker2010}.
However, the absence of Cr segregation at 500~K cannot be explained by the vacancy mechanism alone: rate-theory predictions considering only preferential vacancy-mediated transport still predict a finite Cr depletion at this temperature (Fig.~\ref{fig:pf_NiCr}a).
This discrepancy between experiment and the vacancy-only model is resolved by accounting for preferential SIA-mediated Cr transport.
Cr forms stable mixed $\langle 100 \rangle$ Ni--Cr dumbbells \cite{tucker2010,barnard2012,barnard2014aimd}, and the resulting preferential SIA--Cr drag toward sinks produces an enrichment contribution that offsets vacancy-driven depletion.
As shown in Fig.~\ref{fig:diff_data}b, the difference between the Cr--SIA and Cr--vacancy PDC ratios $(d_{\text{CrI}}/d_{\text{NiI}} - d_{\text{CrV}}/d_{\text{NiV}})$ is nearly zero at 500~K (for the case with preferential SIA-mediated Cr transport included), and the model incorporating both contributions predicts near-zero net segregation, in agreement with experiment.
At 800 and 1000~K, the Cr--SIA binding becomes less effective, and thus, the vacancy mechanism dominates and the model recovers net Cr depletion consistent with experimental measurements (Fig.~\ref{fig:pf_NiCr}).
Inclusion of excess bulk sinks further improves model--experiment agreement at 800~K (Fig.~\ref{fig:pf_NiCr}b), similar to that in Ni--1Fe.
% Unlike Ni--1Fe, no voids were observed in the analyzed Ni--1Cr GB regions at 800~K (Fig.~\ref{fig:exp:NiCr}a), 
% suggesting that the RIS suppression at this temperature is attributable to a mean-field increase in bulk sink 
% density rather than local void competition; this point is discussed further in Sec.~\ref{sec:bulk_sinks}.

The experimental trends are further supported by comparison with prior studies in other Ni--Cr alloys.
Barr et al.~\cite{barr2015} observed Cr depletion in Ni--5Cr under heavy-ion irradiation with magnitude varying with GB character, and Briggs et al.~\cite{briggs2016} reported peak Cr depletion at intermediate temperatures in Ni--18Cr under both proton and Ni-ion irradiation.
The present Ni--1Cr results extend these observations to the dilute limit, where the SCMF transport coefficients are most rigorously applicable \cite{toijer_solute-point_2021}.

The present results can be compared directly with prior ab initio and rate-theory studies of Cr transport in Ni--Cr~\cite{tucker2010,barnard2012,barnard2014aimd}.
Barnard et al.~\cite{barnard2012} showed that RIS predictions for Cr are highly sensitive to the parameterization of SIA transport, with different choices yielding either net Cr enrichment or depletion at GBs.
Barnard and Morgan~\cite{barnard2014aimd} subsequently resolved this ambiguity using ab initio molecular dynamics by finding that more rigorous treatment of SIA dynamics reduced the predicted enrichment contribution and yielded net Cr depletion above $\sim$500~K.
The present dilute Ni--1Cr experimental results confirm this 
hierarchy. 
% vacancy-dominated depletion at 800 and 1000~K with near-zero segregation at 500~K, where SIA transport partially compensates.

As with Fe, the overall modeled temperature dependence of Cr depletion (with default bulk sink strength) follows a bell-shaped trend (Fig.~\ref{fig:pf_NiFe})~\cite{was2007fundamentals,ardell2016}.
However, the additional SIA contribution modifies the low-temperature behavior relative to Fe, suppressing Cr depletion at 500~K.
This distinction shows that even among solutes that deplete via the IK mechanism, the presence or absence of a competing SIA contribution can produce qualitatively different temperature-dependent segregation behavior.
% The model--experiment agreement across three temperatures therefore provides direct support for the competing vacancy and SIA contributions to Cr segregation.
The model--experiment agreement across three temperatures therefore provides direct support for the competing vacancy and SIA contributions to Cr segregation. 
% \sout{, while the remaining discrepancies such as variations in experimental depletion widths point to GB character and local sink geometry effects that the present rate-theory phase-field model does not resolve.}

\subsection{Combined RIS and RETS in Ni--Mn}
\label{sec:disc:NiMn}

Mn exhibits segregation behavior fundamentally distinct from that of Fe and Cr.
The pre-irradiated sample shows measurable Mn enrichment at GBs (Fig.~\ref{fig:exp:NiMn_preirradiation}), indicating thermal segregation established during the initial 1473 K/48~h heat treatment.
The unirradiated region at 500~K similarly exhibits GB enrichment (Fig.~\ref{fig:exp:NiMn}b), confirming that TS occurs independently of irradiation.
The concept of RETS---in which irradiation accelerates the kinetics of approach to thermodynamic equilibrium at GBs---was proposed by Faulkner et al.~\cite{faulkner1996model} and further developed by Nastar~\cite{nastar2005}, but direct experimental evidence in a controlled binary system has been lacking.

The experimental segregation profiles in Ni--1Mn are remarkably similar between the irradiated and unirradiated regions, particularly at 500~K (Fig.~\ref{fig:exp:NiMn}).
At 800 and 1000~K, pronounced profile asymmetry is observed in both the irradiated and unirradiated regions.
Because the unirradiated region contains no irradiation-induced damage and therefore cannot develop asymmetric bulk sink distributions, the presence of asymmetry at 5~$\mu$m depth indicates that GB migration, and not sink asymmetry, is the primary cause of profile asymmetry at these temperatures.
Because the same GB extends continuously through the foil thickness, migration initiated in the irradiated region is also the most likely origin of the coupled GB movement observed in the unirradiated 5~$\mu$m region.
The observed asymmetry at 500~K is weaker but still present, consistent with solute drag at a lower GB velocity.
This interpretation is consistent with recent work highlighting the need to account explicitly for depth, GB diffusion, and surface/end-of-range effects when analyzing RIS under heavy-ion irradiation~\cite{offidani2025analysis}.

The rate-theory phase-field model clarifies the competition between RIS and RETS.
While pre-irradiation TS refers to the segregation established during the 1473~K/48~h heat treatment, RETS denotes the subsequent radiation-enhanced evolution of this pre-existing profile toward the thermodynamically preferred GB segregation state at the irradiation temperature.
Both RETS-only and RIS+RETS simulations were performed for comparison (Fig.~\ref{fig:pf_NiMn}).
Pure RETS simulations predict narrow enrichment peaks confined to a few nanometers around the GB, consistent with the unirradiated-region profiles and with a thermodynamic driving force that remains localized to the interface.
RIS-only simulations predict Mn depletion at all temperatures, reflecting preferential vacancy-mediated Mn transport away from the GB.
The combined RIS+RETS predictions produce non-monotonic profiles with thermodynamic enrichment at the GB core and RIS depletion in the surrounding diffusional tail.
At 500~K, the equilibrium segregation itself is greater than at 1473~K, reflecting the temperature dependence of $\Delta E^{\text{seg}}_{\text{Mn}}$ (Supplementary Material, Sec.~S4); accordingly, RETS dominates the early response and rapidly drives the GB concentration from its initial post-heat-treatment value of approximately 1.5~at.\% toward this higher equilibrium, reaching approximately 3.2~at.\% under RETS-only by 2~dpa. RIS develops over a larger dose and drives Mn depletion, so the coupled RIS+RETS response initially increases, reaches a maximum near 0.15~dpa, and then decreases as RIS increasingly opposes RETS, approaching a quasi-steady GB concentration of approximately 1.9~at.\% by 3--4~dpa rather than the RETS-only equilibrium value (Supplementary Material, Sec.~S6).
% At 500~K, the RIS+RETS prediction is broadly consistent with the experimental irradiated-region profile, although the predicted depletion regions surrounding the enrichment peak are broader than observed experimentally, suggesting that either bulk sinks suppress the RIS contribution more effectively than the mean-field model captures, or that Mn and Ni vacancy diffusivities are more similar than the current parameterization predicts.
Model predictions further show that the non-monotonic profile shape remains stable beyond 10~dpa (Supplementary Material), indicating a genuine steady state rather than a transient low-dose feature.

Two-dimensional bicrystal simulations confirm the depth-dependent transition from RIS+RETS profiles at $\sim$1.5~$\mu$m to TS at 5~$\mu$m (Fig.~\ref{fig:pf_NiMn_2D}), with the similarity between the two regions reflecting the weak net RIS contribution for Mn relative to the thermodynamic enrichment.
W-shaped profiles have been reported in the literature but their mechanistic origin has remained debated.
Barr et al.~\cite{barr2018oscillatory_ris} observed stable W-shaped profiles persisting to 31~dpa in neutron-irradiated 316 stainless steel, contrary to the expectation that RIS would eventually dominate at high dose.
The present simulation results provide a mechanistic basis for this stability: when thermodynamic and kinetic contributions compete, the non-monotonic shape arises as a steady-state feature similar to that observed in simulation results for austenitic Fe-Cr-Ni~\cite{kadambi2025}.
% To the best of our knowledge, this is the first demonstration of such profiles in a dilute binary alloy where the competing mechanisms can be unambiguously isolated.
% We further note that Mn exhibits opposite kinetic behavior in bcc Fe, where dumbbell-mediated transport drives enrichment at sinks rather than depletion~\cite{messina2020,kadambi2025}, underscoring that RIS predictions are crystal-structure specific and must be validated on a system-by-system basis.
The profile asymmetries at 800~K and 1000~K are further addressed in the context of bulk sink effects and GB migration in Sec.~\ref{sec:bulk_sinks} and~\ref{sec:gb_migration}, respectively.

\subsection{Effect of bulk sinks at 800~K}
\label{sec:bulk_sinks}

\begin{sloppypar}
TEM characterization at 800~K reveals the presence of voids in all three alloys, with an average diameter of $\sim$32~nm and a number density of $\sim$1.6~$\times$~10$^{21}$~m$^{-3}$ (Figs.~\ref{fig:exp:NiFe}a and \ref{fig:exp:NiMn}a).
No comparable void population is observed at 500~K or 1000~K, consistent with the characteristic peak swelling temperature behavior in Ni: 500~K falls below the void stability threshold where vacancy mobility is insufficient for nucleation and growth, while 1000~K exhibits enhanced recombination rates and low vacancy supersaturation that suppress void formation \cite{garner2008,chen2023acta,thuinet2018multiscale,was2016emulating}.
EDS maps acquired at 800~K further show measurable Fe and Mn depletion at void surfaces (Figs.~\ref{fig:exp:NiFe}a and \ref{fig:exp:NiMn}a), confirming that voids act not only as point defect sinks but also as sites of solute redistribution, directly competing with GBs for the available solute and defect flux.
Although voids are observed in the Ni--1Cr microstructure at 800~K, none were detected in the immediate vicinity of the analyzed GB regions.
The reduced GB segregation magnitude observed experimentally at 800~K (Figs.~\ref{fig:pf_NiFe}b, \ref{fig:pf_NiCr}b, \ref{fig:pf_NiMn}d) is therefore attributable to this sink competition effect.
\end{sloppypar}

Rate-theory model predictions incorporating excess bulk sinks reproduce the experimentally observed reduction in GB segregation at 800~K for all three alloys (Figs.~\ref{fig:pf_NiFe}b, \ref{fig:pf_NiCr}b, \ref{fig:pf_NiMn}d). This is consistent with literature computational studies showing that increased bulk sink density reduces GB segregation in Ni--Cr \cite{gencturk2023}.
Using the characterized average void radius $r_{\text{void}} \approx 16$~nm and number density $N_{\text{void}} \approx 1.6 \times 10^{21}$~m$^{-3}$, the corresponding mean-field void sink contribution is $k_b^2 \approx 4 \pi r_{\text{void}} N_{\text{void}} \approx 3 \times 10^{14}$~m$^{-2}$.
This value is slightly higher than the default sink strength used in the model ($10^{14}$~m$^{-2}$), but is far below the effective value required to reproduce the experimentally observed suppression of RIS at 800~K ($k_b^2 = 4 \times 10^{17}$~m$^{-2}$).
We interpret this larger effective sink strength as accounting for the large population of bulk defect sinks not captured by TEM-based void characterization, which entails small vacancy clusters and SIA loops below the TEM detection limit, in addition to the resolvable void population. This is consistent with reported loop and void populations in Ni--Cr alloys under comparable irradiation conditions, which correspond to sink strengths of order $10^{14}$--$10^{15}$~m$^{-2}$~\cite{barr2015,briggs2016}, and with positron annihilation spectroscopy studies showing that sub-resolution cluster densities can exceed TEM-visible defect densities by up to two orders of magnitude at comparable dose~\cite{song2024dose}.
Moreover, the solute depletion observed at void surfaces in EDS maps (Figs.~\ref{fig:exp:NiFe}a and \ref{fig:exp:NiMn}a) indicates that voids and loops act as sinks for solute atoms as well as point defects, a mechanism not included in the present mean-field model, which treats bulk sinks as absorbing point defects only.
The elevated sink strength required in the model should therefore be interpreted as an effective parameter that captures the aggregate influence of heterogeneous bulk microstructural sinks.
More quantitative treatment would require explicit modeling of solute redistribution to void and loop surfaces and of the spatial correlation between those sinks and nearby GBs \cite{was2007fundamentals,ardell2016}.

\subsection{Effect of GB migration on segregation}
\label{sec:gb_migration}

Asymmetric segregation profiles are observed in Ni--1Mn at 800~K and 1000~K in the irradiated region (Fig.~\ref{fig:exp:NiMn}a), and a mild asymmetry is also present at 500~K, with Mn increasing toward the GB on one side and dropping sharply on the other.
Such asymmetry is inconsistent with a stationary GB under uniform irradiation conditions. 
% \sour{and suggests either GB migration during irradiation or an asymmetry in bulk sink density between the two adjacent grains.}
Radiation-induced GB migration has been shown in austenitic alloys under both electron and ion irradiation \cite{watanabe1996rigm,barr2022science}, with GB velocities estimated in the range $\sim$0.01--1~nm/s under comparable irradiation conditions \cite{offidani2025}.
The physical origin likely lies in asymmetric point-defect absorption at the GB: molecular dynamics simulations have shown that irradiated GBs are significantly more mobile than unirradiated GBs due to SIA accumulation at the GB, which reduces migration barriers \cite{jin2019scripta,chesser2024prm}.

This behavior is consistent with phase-field moving-GB models that couple RIS with GB migration, which have demonstrated that solute drag and profile asymmetry are sensitive to the relative magnitudes of GB velocity and radiation-enhanced solute diffusivity \cite{rezwan2022effect}.
Heterogeneous bulk sinks, GB character and inclination can also produce asymmetric profiles. However, the presence of asymmetry in the sink-free unirradiated region, its systematic strengthening with temperature, and its characteristic moving-boundary morphology---quantitatively reproduced by the concurrent calculations---together support GB migration as the dominant origin.
Mn is the only solute studied in the present work whose segregation acts as a persistent marker of boundary displacement, likely owing to its favorable thermodynamic GB segregation energy. The absence of asymmetry in Ni--1Fe and Ni--1Cr does not necessarily exclude comparable GB migration in those alloys, and further study is needed to detect it.
The concurrent moving-GB calculations, initialized from the segregation state produced by the 1473~K heat treatment, fit GB velocity independently at each temperature rather than holding it fixed, and reproduce the experimental profile asymmetry at 500 and 800~K (Fig.~\ref{fig:concurrent_postHT_movingGB}). The predicted GB velocities, approximately $10^{-3}$~nm/s at 500~K and $10^{-1}$~nm/s at 800~K, rise nearly two orders of magnitude over this temperature range, consistent with the expected increase in irradiation-enhanced GB mobility with temperature~\cite{jin2019scripta}; since Mn diffusivity also increases with temperature, this larger velocity is required to keep the Péclet number, $\text{Pe}=v\delta/D_{\text{solute}}$ \cite{alkayyali2024actamat,abdeljawad2015}, from falling below unity and the asymmetry from vanishing. The 500~K value lies below, whereas the 800~K value lies within, the range of $\sim$0.01--1~nm/s reported for austenitic alloys under self-ion irradiation at 673~K to 15--70~dpa \cite{offidani2025}. These results support GB migration as a plausible origin of the asymmetric profiles. However, at 1000~K, the present transport parameterization predicts rapid Mn redistribution relative to the tested GB velocity and consequently much weaker asymmetry than observed experimentally. Accurate quantitative prediction at 1000~K requires temperature-dependent parameterization of both the radiation-enhanced Mn diffusivity and GB velocity.

\subsection{Limitations and future directions}
\label{sec:limitations}

% Several experimental limitations should be noted.
% Only random high-angle GBs were sampled, without systematic measurements on special boundaries known to exhibit significantly reduced RIS relative to random boundaries \cite{barr2015} since sink efficiency  
% % introducing scatter that the present 1D mean-field model cannot resolve; this sensitivity to GB structure is also consistent with atomistic studies showing that defect mobility and sink efficiency 
% depends strongly on GB character \cite{uberuaga2015sink}.

Several experimental limitations should be noted.
Only random high-angle GBs were sampled, without systematic measurements on special boundaries. Such boundaries are known to exhibit significantly reduced RIS relative to random boundaries \cite{barr2015}, since sink efficiency depends strongly on GB character \cite{uberuaga2015sink}.
STEM-EDS provides spatial resolution of $\sim$1--2~nm and detection limits of $\sim$0.1~at.\%, which is adequate for the depletion zones observed but potentially underestimates the narrow RETS enrichment peaks predicted for Ni--Mn; atom probe tomography would provide a more quantitative analysis \cite{lach2021stem_apt_ris,carter1994}.
% GB migration is inferred from profile asymmetry alone, and alternative sources of asymmetry --- including GB character anisotropy and grain-to-grain differences in local sink density --- cannot be definitively excluded without pre- and post-irradiation EBSD mapping or in-situ TEM irradiation experiments \cite{barr2022science}.
GB migration is inferred from profile asymmetry alone (Sec.~\ref{sec:gb_migration}); definitive conclusion would require pre- and post-irradiation EBSD mapping or in-situ TEM irradiation experiments \cite{barr2022science}.
Finally, the present study is limited to 2~dpa; higher-dose experiments are needed to attain steady-state evolution and directly test the predicted stability of the segregation profiles.

The rate-theory model also carries several simplifying assumptions.
The mean-field bulk sink strength treatment captures the first-order reduction in GB defect flux at 800~K but neglects spatial correlations between the bulk sinks and GBs, and the segregation to these sinks \cite{was2007fundamentals,ardell2016}.
The Cr--SIA PDC ratio is qualitatively estimated using relative binding energies rather than computed directly via SCMF theory (like that for vacancy PDC ratios); thus, the precise crossover temperature between SIA- and vacancy-dominated Cr transport is uncertain.
The present model does not explicitly treat how unequal vacancy and SIA absorption at GBs modifies local lattice-site balance and thus neglects the defect-flux asymmetry contribution to GB motion; recent rate-theory work further shows that RIS predictions are sensitive to biases in point-defect production and absorption \cite{oplinger2026}.
The injected interstitial effect near the implanted-ion range is also neglected, and future work should incorporate both this effect and GB-character dependence to improve quantitative predictive capability \cite{gaston2016,hoffman2023ion,offidani2025analysis,zinkle2018ion}.
% For Ni--1Mn, the moving-GB simulations use a single velocity ($v \approx 1.7 \times 10^{-2}$~nm/s) at all temperatures, which underestimates the spatial extent of the experimental asymmetry at 800 and 1000~K; temperature-dependent GB velocities are expected to improve quantitative agreement but require independent experimental determination.
Additionally, the predicted RIS depletion regions surrounding the RETS enrichment peak in Ni--1Mn are broader than observed experimentally at 500~K, which may indicate that the Mn--Ni vacancy diffusivity ratio from Toijer et al.~\cite{toijer_solute-point_2021} overestimates the RIS contribution in this system, or that the sub-resolution bulk sinks discussed in Sec.~\ref{sec:bulk_sinks} suppress RIS more effectively at 500~K than the mean-field model captures.
Ballistic mixing effects were generally neglected. At 500~K, simulations performed to test the effect of ballistic relocation (Supplementary Material, Sec.~S7) showed reductions of approximately 5.5\% in the RIS depletion amplitude and 25\% in the RETS enrichment. The effect is expected to be less significant at higher temperatures. Because ballistic mixing does not change the direction of segregation produced by either mechanism, it was neglected in the calculations reported in this work.
Quantitative model--experiment comparisons in the future will require accounting for in-cascade recombination through an effective production efficiency that relates the nominal NRT damage rate to the production rate of freely migrating point defects (see Supplementary Material, Sec.~S8, for a sensitivity analysis).

\section{Conclusions}
\label{sec:conclusions}

In this study, we combined systematic ion irradiation experiments with rate-theory phase-field modeling to deconvolute segregation mechanisms in dilute Ni--1~at.\% Fe, Ni--1~at.\% Cr, and Ni--1~at.\% Mn binary alloys at 500, 800, and 1000~K.
We identified element-specific, temperature-dependent, and microstructure-mediated mechanisms leading to GB segregation under irradiation.
(i) Fe depletes predominantly by preferential vacancy-mediated transport via the IK mechanism across all temperatures, and SCMF transport coefficients parameterized modeling predictions quantitatively reproduce its magnitude.
(ii) Cr depletes through competing vacancy- and SIA-mediated transport: preferential Cr--SIA transport suppresses the net segregation at 500~K, whereas weaker Cr--SIA transport at 800 and 1000~K restores vacancy-dominated depletion.
(iii) Mn exhibits fundamentally distinct behavior dominated by TS, which drives GB enrichment and competes with kinetically driven RIS depletion.
Through depth-dependent characterization, we separated these contributions within the same specimen, with the unirradiated region at 5~$\mu$m reflecting TS and the irradiated region at $\sim$1.5~$\mu$m sampling combined RIS+RETS.
% To the best of our knowledge, this represents the first experimental isolation of competing RIS and RETS contributions in a dilute binary alloy, providing a mechanistic basis for the stable non-monotonic profiles observed in concentrated irradiated alloys.
The coupled model predicts non-monotonic profiles from the RIS--RETS competition that remain stable beyond 10~dpa.

We further found that microstructural effects modify segregation behavior independently of solute identity.
Bulk sinks, including TEM-resolved voids and sub-resolution defect clusters at 800 K, are attributed to a reduction in GB segregation magnitude, with additional reductions caused by RIS-induced solute depletion at void surfaces in all three alloys.
% the effective sink strength required to match experiment exceeds the characterized void population, indicating additional contributions from sub-resolution defect clusters, evidenced directly by Fe and Mn depletion at void surfaces in EDS maps.
% Asymmetric Mn segregation profiles all three temperatures in both irradiated and unirradiated regions are consistent with irradiation-induced GB migration, and are demonstrated by moving-GB simulations.
Asymmetric Mn segregation profiles in both irradiated and unirradiated regions are consistent with irradiation-induced GB migration, with the moving-GB simulations largely supporting this interpretation.
% and moving-GB simulations reproduce the opposing asymmetry patterns produced by RETS and RIS+RETS relative to the direction of GB motion.
Together, our combined experimental and modeling results demonstrate that predictive modeling of GB chemistry evolution under irradiation requires concurrent treatment of defect-mediated kinetic transport, thermodynamic driving forces, and microstructural sink effects.
The dilute binary alloy framework established here provides a mechanistic foundation for extending segregation predictions to multicomponent Ni-based reactor alloys and informing alloy design for advanced nuclear applications.

\section*{ACKNOWLEDGMENTS}

This work was supported by the Laboratory-Directed Research and Development (LDRD) program of Idaho National Laboratory under Project No. 23A1070-092FP.
This research made use of Idaho National Laboratory's High Performance Computing systems located at the Collaborative Computing Center and supported by the Office of Nuclear Energy of the U.S.
Department of Energy and the Nuclear Science User Facilities under Contract No.
DE-AC07-05ID14517.

\section*{CRediT authorship contribution statement}

\textbf{Sohail Shah:} Conceptualization, Validation, Formal analysis, Investigation, Resources, Data curation, Writing -- original draft, Writing -- review \& editing, Visualization.
\textbf{Sourabh Bhagwan Kadambi:} Conceptualization, Methodology, Software, Formal analysis, Writing -- original draft, Writing -- review \& editing, Visualization, Funding acquisition.
\textbf{Anshul Kamboj:} Data curation, Formal analysis, Validation, Investigation.
\textbf{Benjamin Diaz:} Investigation, Formal analysis, Visualization.
\textbf{Xinchang Zhang:} Investigation.
\textbf{Jana Howard:} Investigation.
\textbf{Michael Moorehead:} Investigation.
\textbf{Boopathy Kombaiah:} Conceptualization, Validation, Resources, Supervision, Funding acquisition.
\textbf{Lin Shao:} Resources, Data curation, Supervision.
\textbf{Mukesh Bachhav:} Conceptualization, Validation, Resources, Writing -- review \& editing, Supervision, Project administration, Funding acquisition.

\section*{Declaration of generative AI and AI-assisted 
technologies in the writing process}

During the preparation of this work, the authors used Claude Sonnet 5.5 (Anthropic) and GPT-5.6 through ChatGPT (OpenAI) to improve sentence phrasing and the readability of the manuscript. After using these tools, the authors reviewed and edited the content as needed and take full responsibility for the content of the publication.

\bibliography{ref}
\end{document}